\documentclass[12pt]{elsart}
\usepackage{amsfonts}
\usepackage{natbib}
\usepackage{colonequals} 
\usepackage{amsmath}
\usepackage{amssymb,epsfig,calc}

\usepackage[latin1]{inputenc} 
\usepackage{slashbox}
\usepackage{natbib}
\usepackage{amsmath}
\usepackage{amsfonts}
\usepackage{amssymb}
\usepackage{ctable}
\usepackage{subfigure}
\usepackage{graphicx}
\usepackage{cancel}
\usepackage{array}
\usepackage{color}
\usepackage{colortbl} 
\usepackage{multirow} 
\usepackage{subfigure} 
\usepackage{latexsym}
\usepackage{textcomp}
\usepackage{tipa}
\usepackage{pstricks,pst-node,pst-tree,pstricks-add}
\usepackage{url}

\makeatletter
\renewcommand*\env@matrix[1][*\c@MaxMatrixCols c]{%
  \hskip -\arraycolsep
  \let\@ifnextchar\new@ifnextchar
  \array{#1}}
\makeatother

\def\real{\mathbb{R}}

\def\be{\boldsymbol{e}}
\def\bx{\boldsymbol{x}}

\def\bz{\boldsymbol{z}}

\def\bm{\boldsymbol{m}}

\def\bu{\boldsymbol{u}}

\def\bA{\boldsymbol{A}}
\def\bB{\boldsymbol{B}}
\def\bC{\boldsymbol{C}}
\def\bD{\boldsymbol{D}}

\def\bX{\boldsymbol{X}}

\def\bS{\boldsymbol{S}}
\def\bI{\boldsymbol{I}}

\def\bU{\boldsymbol{U}}

\def\bZ{\boldsymbol{Z}}
\def\bzero{\boldsymbol{0}}
\def\b1{\boldsymbol{1}}

\def\bmu{\mbox{\boldmath $\mu$}}

\def\bbeta{\mbox{\boldmath $\beta$}}

\def\bgamma{\mbox{\boldmath $\gamma$}}

\def\btheta{\mbox{\boldmath $\theta$}}

\def\bSigma{\mbox{\boldmath $\Sigma$}}
\def\bGamma{\mbox{\boldmath $\Gamma$}}
\def\bLambda{\mbox{\boldmath $\Lambda$}}

\def\bPsi{\mbox{\boldmath $\Psi$}}
\def\bTheta{\mbox{\boldmath $\Theta$}}

\def\Cov{\text{Cov}}
\def\Var{\text{Var}}
\def\diag{\text{diag}}
\def\tr{\text{tr}}

\begin{document}

\begin{frontmatter}
\title{Clustering and Classification via Cluster-Weighted Factor Analyzers}
\author[UG]{Sanjeena Subedi},
\ead{ssubedi@uoguelph.ca}
\author[UC]{Antonio Punzo},
\ead{antonio.punzo@unict.it}
\author[UC]{Salvatore Ingrassia\corauthref{cor}},
\ead{s.ingrassia@unict.it}
\author[UG]{Paul D. McNicholas}
\ead{paul.mcnicholas@uoguelph.ca}
\corauth[cor]{Corresponding author}
\address[UG]{Department of Mathematics and Statistics, University of Guelph \\  Ontario, Canada, N1G~2W1.}
\address[UC]{Department of Economics and Business, University of Catania \\ Corso Italia 55, 95129 Catania, Italy.}

\begin{abstract}
In model-based clustering and classification, the cluster-weighted model constitutes a convenient approach when the random vector of interest constitutes a response variable $Y$ and a set $p$ of explanatory variables $\bX$. However, its applicability may be limited when $p$ is high. To overcome this problem, this paper assumes a latent factor structure for $\bX$ in each mixture component. This leads to the cluster-weighted factor analyzers (CWFA) model. By imposing constraints on the variance of $Y$ and the covariance matrix of $\bX$, a novel family of sixteen CWFA models is introduced for model-based clustering and classification. The alternating expectation-conditional maximization algorithm, for maximum likelihood estimation of the parameters of all the models in the family, is described; to initialize the algorithm, a 5-step hierarchical procedure is proposed, which uses the nested structures of the models within the family and thus guarantees the natural ranking among the sixteen likelihoods. Artificial and real data show that these models have very good clustering and classification performance and that the algorithm is able to recover the parameters very well.  
\end{abstract}
\end{frontmatter}

\section{Introduction}
\label{sec:intro}

In direct applications of finite mixture models, each of the $G$ mixture components is taken to represent a sub-group (or cluster) within the data \citep[see][pp.~2--3]{Titt:Smit:Mako:stat:1985}. The terms `model-based clustering' and `model-based classification' have been used to describe the adoption of mixture models or, more often, a family of mixture models for clustering and classification, respectively. 
In the 1990's, three model-based clustering papers (\citealp{Banf:Raft:mode:1993}, \citealp{Cele:Gova:Gaus:1995}, and \citealp{Ghah:Hint:TheE:1997}) effectively set the scene for the push towards the finite mixture model-based approaches that followed. Overviews of mixture models and their applications are given in \citet{EvHa:fini:1981}, \citet{Titt:Smit:Mako:stat:1985}, \citet{McLa:Peel:fini:2000}, and \citet{Fruh:Fine:2006}.

Consider a random vector $\left(\bX',Y\right)'$, defined from $\Omega$ to $\real^p\times \real$, where a latent group-structure as well as a linear dependence of $Y$ on $\bx$ in each group are assumed. 
Under these assumptions, the linear cluster-weighted model \citep[CWM; introduced in ][]{Gers:Nonl:1997} is an ideal choice within the mixture modelling framework. 
It factorizes the joint density of $\left(\bX',Y\right)'$, in each mixture-component, into the product of the conditional density of $Y|\bx$ and the marginal density of $\bX$. In this manner, the model takes into account the potential of finite mixtures of regressions \citep[see][Chapter~8]{Fruh:Fine:2006} in modelling the conditional density of $Y|\bx$, and the potential of finite mixtures of Gaussian distributions (see \citealp{Titt:Smit:Mako:stat:1985} and \citealp{McLa:Peel:fini:2000}) in modelling the joint density of $\left(\bX',Y\right)'$ and the marginal density of $\bX$.

Recent  literature on model-based clustering and classification through the CWM can be summarized as follows.
\citet*{Ingr:Mino:Vitt:Loca:2012} study the relationships between the linear Gaussian CWM and some well-known mixture-based approaches, moreover they
 consider the $t$-distribution as a  robust alternative to  Gaussian assumptions. 
By using this model as a building block, \citet*{Ingr:Mino:Punz:Mode:2012} introduce a family of parsimonious linear $t$-CWMs for model-based clustering. Finally, under Gaussian assumptions for both mixture-component densities, \citet{Punz:Flex:2012} introduces the polynomial CWM as a flexible tool for clustering and classification purposes

However, the applicability of linear Gaussian CWMs in high dimensional $\bX$-spaces still remains a challenge.
The number of parameters for this model is $\left(G-1\right)+G\left(p+2\right)+G\left[p+p\left(p+1\right)/2\right]$, of which $Gp\left(p+1\right)/2$ are used for the group covariance matrices $\bSigma_g$ of $\bX$ alone, $g=1,\ldots,G$, and this increases quadratically with $p$.
To overcome this issue, we assume a latent Gaussian factor structure for $\bX$, in each mixture-component, which leads to the {\em Factor Regression Model} (FRM) of $Y$ on $\bx$ (see \citealp{West:Baye:2003}, \citealp{Wang:Carv:Luca:West:Bull:2007}, and \citealp{Carv:Chan:Luca:Nevi:Wang:West:High:2008}). 
The FRM assumes $\bSigma_g=\bLambda_g\bLambda_g'+\bPsi_g$, where the loading matrix is a $p\times q$ matrix of parameters typically with $q\ll p$ and the noise matrix $\bPsi_g$ is a diagonal matrix.  
The adoption of this group covariance structure in the linear Gaussian CWM framework leads to the linear Gaussian cluster-weighted factor analyzers model (CWFA), which is characterized by $G\left[pq-q\left(q-1\right)/2\right]+Gp$ parameters for the group covariance matrices.
The CWFA model follows the principle of the general form of mixtures of factor analyzers regarding $\bX$; mixtures of factor analyzers were developed by \citet[][Chapter~8]{McLa:Peel:fini:2000} and \citet{McLa:Peel:Bean:Mode:2003} on the basis of the original work of \cite{Ghah:Hint:TheE:1997}. 
Furthermore, starting from the works of \citet{McNi:Murp:Pars:2008}, \citet{McNi:Mode:2010}, \citet*{Ingr:Mino:Vitt:Loca:2012} and \citet*{Ingr:Mino:Punz:Mode:2012}, a novel family of sixteen mixture models --- obtained as special cases of the linear Gaussian CWFA by conveniently constraining the component variances of $Y$ and $\bX$ --- is introduced to facilitate parsimonious model-based clustering and classification in the defined paradigm.

The paper is organized as follows.
Section~\ref{sec:Model genesis} recalls the linear Gaussian CWM and the FRM (with details given in \appendixname~\ref{app:Conditional distribution of Y|X,u}); they are the basic models to define the linear Gaussian CWFA models introduced in Section~\ref{The modelling framework}.
Model fitting with the alternating expectation-conditional maximization (AECM) algorithm is presented in Section~\ref{sec:Estimation via the AECM algorithm}, with details given in \appendixname~\ref{subsec:Constraints on the X variable}.
Section~\ref{sec:Model selection and performance assessment} addresses computational details on some aspects of the AECM algorithm and discusses model selection and evaluation. 
Artificial and real data are considered in Section~\ref{sec:Data analyses}, and the paper concludes with discussion and suggestions for further work in Section~\ref{sec:Conclusions, discussion, and future works}.  

\section{Model}
\label{sec:Model genesis} 

This section provides a step-by-step introduction to the model we introduce in the next section. 

\subsection{The linear Gaussian cluster-weighted model}
\label{subsec:The linear Gaussian Cluster-Weighted Model} 

Let $p\left(\bx,y\right)$ be the joint density of $\left(\bX',Y\right)'$.
Suppose that $\Omega$ can be partitioned into $G$ groups, say $\Omega_1, \ldots,\Omega_G$.
The CWM defines the joint density as
\begin{equation}
p\left(\bx,y;\btheta\right)=\sum^{G}_{g=1}\pi_g p\left(y|\bx,\Omega_g\right)p\left(\bx|\Omega_g\right), \label{eq:CWM base} 
\end{equation}
where $p\left(y|\bx,\Omega_g\right)$ is the conditional density of the response variable $Y$ given $\bx$ and $\Omega_g$, $p(\bx|\Omega_g)$ is the marginal density of $\bx$ given $\Omega_g$, $\pi_g=p(\Omega_g)$ is the weight of $\Omega_g$ in the mixture (defined so that $\pi_g > 0$ and $\sum_g \pi_g = 1$), $g=1,\ldots,G$, and $\btheta$ contains all of the parameters in the mixture. 

The component densities $p\left(\bx|\Omega_g\right)$ and $p\left(y|\bx,\Omega_g\right)$ are usually assumed to be (multivariate) Gaussian (see, e.g., \citealp*{Ingr:Mino:Vitt:Loca:2012} and \citealp{Punz:Flex:2012}), the former with mean vector $\bmu_g$ and covariance matrix $\bSigma_g$ and the latter with linear conditional mean $\mu\left(\bx,\bbeta_g\right)=\beta_{0g} + \bbeta'_{1g} \bx$ and conditional variance $\sigma^2_g$, where $\bbeta_g=\left(\beta_{0g},\bbeta_{1g}'\right)'$, $\beta_{0g} \in \real$, and $\bbeta_{1g}\in \real^p$.
In other words, conditional on $\bx$ and $\Omega_g$, the linear model $Y|\bx=\mu\left(\bx,\bbeta_g\right)+\varepsilon_g$ holds.
Thus, the general CWM in \eqref{eq:CWM base} becomes the linear Gaussian CWM	
\begin{equation}
p\left(\bx,y; \btheta\right)=\sum_{g=1}^G\pi_g \phi\left(y|\boldsymbol{x};\mu\left(\boldsymbol{x};\bbeta_g\right),\sigma^2_g\right)\phi\left(\bx;\boldsymbol{\mu}_g,\bSigma_g\right),
\label{eq:linear Gaussian CWM}
\end{equation}
where
\begin{equation*}\begin{split}
\phi\left(y|\boldsymbol{x};\mu\left(\boldsymbol{x};\bbeta_g\right),\sigma^2_g\right) &= \frac{1}{\sqrt{2 \pi \sigma^2_g}} \exp \left\{ - \frac{\left[y-\mu\left(\boldsymbol{x};\bbeta_g\right)\right]^2}{2\sigma_g^2} \right\},\\
\phi\left(\bx;\bmu_g,\bSigma_g\right) &= \frac{1}{\left(2 \pi\right)^{p/2} |\bSigma_g|^{p/2}} \exp \left\{-\frac{1}{2} \left(\bx - \bmu_g\right)' \bSigma_g^{-1} \left(\bx - \bmu_g\right) \right\}. 
\end{split}\end{equation*}

\subsection{The factor regression model}
\label{subsec:The Factor Regression Model}

The factor analysis model (\citealp{Spea:Thep:1904} and \citealp{Bart:Fact:1953}), for the $p$-dimensional variable $\bX$, postulates that
\begin{equation}
\bX = \bmu + \bLambda \bU + \be,
\label{eq:factor analysis model}
\end{equation}
where $\bU \sim N_q\left(\bzero, \bI_q\right)$ is a $q$-dimensional $(q\ll p)$ vector of latent factors, $\bLambda$ is a $p\times q$ matrix of factor loadings, and $\be \sim N_p\left(\bzero, \bPsi\right)$, with $\bPsi = \diag\left(\psi_1^2, \ldots, \psi_p^2\right)$, independent of $\bU$. 
Then $\bX \sim N_p(\bmu, \bLambda\bLambda'+\bPsi)$ and, conditional on $\bu$, results in $\bX|\bu \sim N_p\left(\bmu+\bLambda\bu,\bPsi\right)$. 

Model \eqref{eq:factor analysis model} can be considered similarly to the standard (linear) regression model $Y=\beta_0+\bbeta_1'\bX+ \varepsilon$ leading to the FRM (see \citealp{West:Baye:2003}, \citealp{Wang:Carv:Luca:West:Bull:2007}, and \citealp{Carv:Chan:Luca:Nevi:Wang:West:High:2008}) 
$$Y = \beta_0+\bbeta'_1(\bmu + \bLambda \bU + \be) + \varepsilon
  = \left(\beta_0 + \bbeta'_1 \bmu\right) + \bbeta'_1 \bLambda \bU + \left(\bbeta'_1 \be + \varepsilon\right),$$
where $\varepsilon$ is assumed to be independent of $\bU$ and $\be$.
The mean and variance of $Y$ are given by
\begin{equation*}\begin{split}
\mathbb{E}\left(Y\right) &= \beta_0 + \bbeta'_1 \bmu \\
\Var\left(Y\right) &= \Var\left(\bbeta'_1 \bLambda \bU\right) + \Var\left(\bbeta'_1 \be\right) + \Var\left(\varepsilon\right)\\ &= \bbeta'_1 \bLambda \bLambda' \bbeta_1 + \bbeta' \bPsi \bbeta_1 + \sigma^2
= \bbeta'_1 \left(\bLambda \bLambda' +\Psi\right) \bbeta_1 + \sigma^2,
\end{split}\end{equation*}
respectively, and so $Y \sim N\left(\beta_0 + \bbeta'_1 \bmu, \bbeta'_1 \left(\bLambda \bLambda' + \Psi\right) \bbeta_1 + \sigma^2\right)$. 

Consider the triplet $\left(Y,\bX',\bU'\right)'$. 
Its mean is given by
\[ 
\mathbb{E}
\begin{bmatrix} 
Y   \\ 
\bX \\ 
\bU  
\end{bmatrix} = \begin{bmatrix} \beta_0 + \bbeta'_1 \bmu \\ \bmu \\ \bzero  \end{bmatrix},
\]
and because $\Cov(\bX,Y)= (\bLambda \bLambda'+\bPsi)\bbeta_1$ and $\Cov(\bU,Y) =  \bLambda' \bbeta_1$, it results
\[
\Cov\begin{bmatrix} 
Y   \\ 
\bX \\ 
\bU  
\end{bmatrix} 
= 
\begin{bmatrix}  
\bbeta'_1 \bSigma \bbeta_1+ \sigma^2&\bbeta'_1\bSigma &\bbeta'_1 \bLambda\\
\bSigma\bbeta_1 & \bSigma & \bLambda\\
\bLambda' \bbeta_1 & \bLambda' & \bI_q \\
\end{bmatrix}, \label{Cov(Y,X,U)}
\]
where $\bSigma=\bLambda \bLambda' + \bPsi$. 
Now, we can write the joint density of $\left(Y,\bX',\bU'\right)'$ as 
\begin{equation}
p\left(y,\bx,\bu\right) = \phi\left(y|\bx,\bu\right)\phi\left(\bx|\bu\right)\phi\left(\bu\right).
\label{eq:joint distribution of the triplet}
\end{equation}
Here, the distribution and related parameters for both $\bX|\bu$ and $\bU$ are known.
Thus, we need only to analyze the distribution of $Y|\bx,\bu$.
Importantly, $\mathbb{E}\left(Y|\bx,\bu\right)= \mathbb{E}\left(Y|\bx\right)$ and $\Var\left(Y|\bx,\bu\right)= \Var(Y|\bx)$, and so $Y|\bx,\bu \sim N\left(\beta_0 + \bbeta'_1 \bx, \sigma^2\right)$; mathematical details are given in \appendixname~\ref{app:Conditional distribution of Y|X,u}. 
This implies that $\phi\left(y|\bx,\bu\right)=\phi\left(y|\bx\right)$ and, therefore, $Y$ is conditionally independent of $\bU$ given $\bX=\bx$, so that \eqref{eq:joint distribution of the triplet} becomes
\begin{equation}
p\left(y,\bx,\bu\right) =  \phi\left(y|\bx\right)\phi\left(\bx|\bu\right)\phi\left(\bu\right).
\label{eq:compacted joint distribution of the triplet}
\end{equation}
Similarly, $\bU|y,\bx \sim N\left(\bgamma \left(\bx-\bmu\right),\bI_q-\bgamma\bLambda\right)$, where $\bgamma=\bLambda'\left(\bLambda\bLambda'+\bPsi\right)^{-1}$, and thus $\bU$ is conditionally independent on $Y$ given $\bX=\bx$. 
Therefore, 
\begin{equation*}\begin{split}
&\mathbb{E}\left[\bU|\bx;\bmu,\bLambda,\bPsi\right]=\bgamma\left(\bx-\bmu\right), \quad \text{and}\\ 
&\mathbb{E}\left[\bU \bU'|\bx;\bmu,\bLambda,\bPsi\right]=\bI_q-\bgamma\bLambda+\bgamma\left(\bx-\bmu\right)\left(\bx-\bmu\right)'\bgamma'. 
\label{eq:U|xy}
\end{split}\end{equation*}

\section{The modelling framework}\label{The modelling framework}

\subsection{The general model}
\label{sec:The General Model}

Assume that for each $\Omega_g$, $g=1, \ldots, G$, the pair $\left(\bX',Y\right)'$ satisfies a FRM, that is
\begin{equation}
Y = \beta_{0g}+\bbeta_{1g}'\bX+ \varepsilon_g  \quad \text{with} \quad  \bX = \bmu_g + \bLambda_g \bU_g + \be_g, 
\label{eq:CWFA model}
\end{equation}
where $\boldsymbol{\Lambda}_g$ is a $p\times q$ matrix of factor loadings, $\bU_g \sim N_q\left(\boldsymbol{0},\bI_q\right)$ is the vector of factors,
$\be_g\sim N_p\left(\boldsymbol{0},\boldsymbol{\Psi}_g\right)$ are the errors, $\boldsymbol{\Psi}_g=\diag\left(\psi_{1g},\ldots,\psi_{pg}\right)$, and $\varepsilon_g \sim N(0,\sigma^2_g)$.
Then the linear Gaussian CWM in \eqref{eq:linear Gaussian CWM} can be extended in order to include the underlying factor structure \eqref{eq:CWFA model} for the $\bX$ variable.
In particular, by recalling that $Y$ is conditionally independent of $\bU$ given $\bX=\bx$ in the generic $\Omega_g$, we get
\begin{equation}
p\left(\bx,y;\btheta\right)
=
\sum_{g=1}^{G} \pi_g\phi\left(y|\bx;\mu\left(\boldsymbol{x};\bbeta_g\right),\sigma^2_g\right)\phi\left(\bx;\bmu_g,\bLambda_g \bLambda_g' + \bPsi_g\right), 
\label{eq:linear Gaussian CWFA}
\end{equation}
where $\btheta=\left\{\pi_g,\bbeta_g,\sigma^2_g,\bmu_g,\bLambda_g,\bPsi_g;g=1,\ldots,G\right\}$. 
Model \eqref{eq:linear Gaussian CWFA} is the linear Gaussian CWFA, which we shall refer to as the CWFA model herein.

\subsection{Parsimonious versions of the model}
\label{sec:Parsimonious versions of the model}

In this section, we extend the linear Gaussian CWFA by allowing constraints across groups on $\sigma^2_g$, $\bLambda_g$, and $\bPsi_g$, and on whether or not $\bPsi_g=\psi_g\boldsymbol{I}_p$ (isotropic assumption).
The full range of possible constraints provides a family of sixteen different parsimonious CWFAs, which are given
in \tablename~\ref{tab:parsimonious_model}.
\begin{table}[!ht]
\begin{center}
\renewcommand{\arraystretch}{1.4}
\caption{Parsimonious covariance structures derived from the CWFA model.}
\label{tab:parsimonious_model}
\resizebox{\textwidth}{!}{
\begin{tabular}{cccccc} 
\toprule
Model ID & $Y$ Variance & Loading Matrix & Error Variance & Isotropic & Covariance parameters \\ 
\midrule
UUUU & unconstrained   & unconstrained   & unconstrained & unconstrained & $G + G\left[pq - q\left(q -1\right)/2\right] + Gp$ \\
UUUC & unconstrained   & unconstrained   & unconstrained & constrained & $G + G\left[pq -q\left(q-1\right)/2\right] +G$\\
UUCU & unconstrained   & unconstrained   & constrained & unconstrained & $G + G\left[pq -q\left(q -1\right)/2\right] + p$ \\
UUCC & unconstrained   & unconstrained   & constrained & constrained & $G + G\left[pq -q\left(q-1\right)/2\right] + 1$ \\
UCUU & unconstrained   & constrained   & unconstrained & unconstrained & $G + \left[pq -q\left(q - 1\right)/2\right] + Gp$ \\
UCUC & unconstrained   & constrained   & unconstrained & constrained & $G + \left[pq -q\left(q - 1\right)/2\right] + G$ \\
UCCU & unconstrained   & constrained   & constrained & unconstrained & $G + \left[pq -q\left(q - 1\right)/2\right] + p$ \\
UCCC & unconstrained   & constrained   & constrained & constrained & $G + \left[pq -q\left(q - 1\right)/2\right] + 1$   \\
CUUU & constrained   & unconstrained   & unconstrained & unconstrained & $1 + G\left[pq - q\left(q -1\right)/2\right] + Gp$ \\
CUUC & constrained   & unconstrained   & unconstrained & constrained & $1 + G\left[pq -q\left(q-1\right)/2\right] +G$\\
CUCU & constrained   & unconstrained   & constrained & unconstrained & $1 + G\left[pq -q\left(q -1\right)/2\right] + p$ \\
CUCC & constrained   & unconstrained   & constrained & constrained & $1 + G\left[pq -q\left(q-1\right)/2\right] + 1$ \\
CCUU & constrained   & constrained   & unconstrained & unconstrained & $1 + \left[pq -q\left(q - 1\right)/2\right] + Gp$ \\
CCUC & constrained   & constrained   & unconstrained & constrained & $1 + \left[pq -q\left(q - 1\right)/2\right] + G$ \\
CCCU & constrained   & constrained   & constrained & unconstrained & $1 + \left[pq -q\left(q - 1\right)/2\right] + p$ \\
CCCC & constrained   & constrained   & constrained & constrained & $1 + \left[pq -q\left(q - 1\right)/2\right] + 1$   \\
\bottomrule
\end{tabular}
}  
\end{center}
\end{table}

Here, models are identified by a sequence of four letters.
The letters refer to whether or not the constraints $\sigma^2_g=\sigma^2$, $\bLambda_g=\bLambda$, $\bPsi_g=\bPsi$, and $\bPsi_g=\psi_g\boldsymbol{I}_p$, respectively, are imposed.
The constraints on the group covariances of $\bX$ are in the spirit of \citet{McNi:Murp:Pars:2008}, while that on the group variances of $Y$ are borrowed from \citet*{Ingr:Mino:Punz:Mode:2012}. 
Each letter can be either C, if the corresponding constraint is applied, or U if the particularconstraint is not applied.  
For example, model CUUC assumes equal $Y$ variances between groups, unequal loading matrices, and unequal, but isotropic, noise.   

\subsection{Model-based classification}
\label{sec:CWFAsemisupervised}

Suppose that $m$ of the $n$ observations in $\mathcal{S}$ are labeled.
Within the model-based classification framework, we use all of the $n$ observations to estimate the parameters in \eqref{eq:linear Gaussian CWFA}; the fitted model classifies each of the $n-m$ unlabeled observations through the corresponding maximum \textit{a posteriori} probability (MAP). 
As a special case, if $m=0$, we obtain the clustering scenario. 
Drawing on \citet{Hosm:acom:1973}, \citet[][Section~4.3.3]{Titt:Smit:Mako:stat:1985} point out that knowing the label of just a small proportion of observations \textit{a priori} can lead to improved clustering performance.   

Notationally, if the $i$th observation is labeled, denote with $\widetilde{\boldsymbol{z}}_i=\left(\widetilde{z}_{i1},\ldots,\widetilde{z}_{iG}\right)$ its component membership indicator. 
Then, arranging the data so that the first $m$ observations are labeled, the complete-data likelihood becomes
\begin{equation*}\begin{split}
L_c\left(\btheta\right) &=
\prod_{i=1}^m\prod_{g=1}^G\left[\pi_g \phi\left(y_i|\boldsymbol{x}_i;\mu\left(\boldsymbol{x};\bbeta_g\right),\sigma^2_g\right) \phi\left(\boldsymbol{x}_i|\bu_i;\bmu_g,\boldsymbol{\Lambda}_g,\boldsymbol{\Psi}_g\right)\phi\left(\bu_{ig}\right)\right]^{\widetilde{z}_{ig}} 
\\ 
& \times
\prod_{i=m+1}^n\prod_{g=1}^G\left[\pi_g \phi\left(y_i|\boldsymbol{x}_i;\mu\left(\boldsymbol{x};\bbeta_g\right),\sigma^2_g\right) \phi\left(\boldsymbol{x}_i|\bu_i;\bmu_g,\boldsymbol{\Lambda}_g,\boldsymbol{\Psi}_g\right)\phi\left(\bu_{ig}\right)\right]^{z_{ig}}. 
\end{split}\end{equation*}
For notational convenience, in this paper we prefer to present the AECM algorithm in the model-based clustering paradigm (cf.\ Section~\ref{sec:Estimation via the AECM algorithm}).
However, the extension to the model-based classification context is simply obtained by substituting the `dynamic' (with respect to the iterations of the algorithm) $\bz_1,\ldots,\bz_m$ with the ``static'' $\widetilde{\bz}_1,\ldots,\widetilde{\bz}_m$.   

\section{Parameter Estimation} 
\label{sec:Estimation via the AECM algorithm}

\subsection{The AECM algorithm}
The AECM algorithm \citep{Meng:VanD:TheE:1997} is used for fitting all the models within the family defined in Section~\ref{tab:parsimonious_model}. 
This algorithm is an extension of the expectation-maximization (EM) algorithm \citep{Demp:Lair:Rubi:Maxi:1977} that uses different specifications of missing data at each stage.
Let $\mathcal{S}=\left\{\left(\bx_i',y_i\right)';i=1,\ldots,n\right\}$ be a sample of size $n$ from \eqref{eq:linear Gaussian CWFA}.
In the EM framework, the generic observation $\left(\bx_i',y_i\right)'$ is viewed as being incomplete; its complete counterpart is given by $\left(\bx_i',y_i,\bu_{ig}',\boldsymbol{z}_i'\right)'$, where $\boldsymbol{z}_i$ is the component-label vector in which $z_{ig}=1$ if $\left(\bx_i',y_i\right)'$ comes from $\Omega_g$ and $z_{ig}=0$ otherwise.
Then the complete-data likelihood, by considering the result in \eqref{eq:compacted joint distribution of the triplet}, can be written as
\begin{displaymath}
L_c\left(\btheta\right) = \prod_{i=1}^n\prod_{g=1}^G\left[\pi_g\phi\left(y_i|\boldsymbol{x}_i; \mu\left(\boldsymbol{x};\bbeta_g\right),\sigma^2_g\right) \phi\left(\bx_i|\bu_i;\bmu_g,\bLambda_g \bLambda_g' + \bPsi_g\right)\phi\left(\bu_{ig}\right)\right]^{z_{ig}}. 
\end{displaymath}

The idea of the AECM algorithm is to partition $\btheta$, say $\btheta_1=\left(\btheta'_1,\btheta'_2\right)'$, in such a way that the likelihood is easy to maximize for $\btheta_1$ given $\btheta_2$ and \textit{vice versa}. The AECM algorithm consists of two cycles, each containing an E-step and a CM-step. The two CM-steps correspond to the partition of $\btheta$ into $\btheta_1$ and $\btheta_2$. Then, we can iterate between these two conditional maximizations until convergence. In the next two sections, we illustrate the two cycles for the UUUU model only.
Details on the other models of the family are given in \appendixname~\ref{app:details on the EM algorithm}.

\subsection{First cycle} 

Here, $\btheta_1=\left\{\pi_g,\bbeta_g,\bmu_g,\sigma^2_g;g=1, \ldots, G\right\}$, where the missing data are the unobserved group labels $\boldsymbol{z}_i$, $i=1,\ldots,n$. 
The complete-data likelihood is
\begin{displaymath}
L_1\left(\btheta_1\right) =\prod_{i=1}^n\prod_{g=1}^G\left[\pi_g \phi\left(y_i|\boldsymbol{x}_i; \mu\left(\boldsymbol{x}_i;\bbeta_g\right),\sigma^2_g\right) \phi\left(\boldsymbol{x}_i;\bmu_g,\bSigma_g\right)\right]^{z_{ig}}. 
\end{displaymath}
Consider the complete-data log-likelihood  
\begin{equation*}\begin{split}
l_{c1}&\left(\btheta_1\right) = \sum_{i=1}^n \sum_{g=1}^G z_{ig} \ln \left[\pi_g\phi \left(y_i|\boldsymbol{x}_i; \mu\left(\boldsymbol{x}_i;\bbeta_g\right),\sigma^2_g\right) \phi\left(\boldsymbol{x}_i;\bmu_g,\boldsymbol{\Lambda}_g,\boldsymbol{\Psi}_g\right) \right] \\
&= - \frac{n\left(p+1\right)}{2} \ln 2\pi - \frac{1}{2} \sum_{i=1}^n \sum_{g=1}^G z_{ig} \ln \sigma^2_g -\frac{1}{2}\sum_{i=1}^n\sum_{g=1}^G z_{ig} \frac{\left(y_i-\beta_{0g}-\bbeta'_{1g} \bx_i\right)^2}{\sigma_g^2} + \\ 
& -\frac{1}{2}\sum_{i=1}^n \sum_{g=1}^G z_{ig} \ln \left|\bSigma_g\right|  -\frac{1}{2}\sum_{i=1}^n\sum_{g=1}^G z_{ig} \left(\bx_i - \bmu_g\right)' \bSigma_g^{-1} \left(\bx_i - \bmu_g\right)+\sum_{g=1}^G n_g \ln \pi_g,
\end{split}\end{equation*}
where $n_g = \displaystyle\sum_{i=1}^n z_{ig}$. 
Because $\bSigma_g = \bLambda \bLambda'_g + \bPsi_g$, we get
\begin{equation*}\begin{split}
l_{c1}&\left(\btheta_1\right) = -\frac{n\left(p+1\right)}{2} \ln 2\pi - \frac{1}{2} \sum_{i=1}^n \sum_{g=1}^G z_{ig} \ln \sigma^2_g\\ 
& -\frac{1}{2}\sum_{i=1}^n\sum_{g=1}^G z_{ig} \frac{\left(y_i-\beta_{0g}-\bbeta'_{1g} \bx_i\right)^2}{ \sigma_g^2} -\frac{1}{2}\sum_{i=1}^n \sum_{g=1}^G z_{ig} \ln \left|\bLambda \bLambda'_g + \bPsi_g\right|\\ 
&  -\frac{1}{2} \sum_{i=1}^n\sum_{g=1}^G z_{ig}\tr\left\{ \left(\bx_i - \bmu_g\right) \left(\bx_i - \bmu_g\right)' \left(\bLambda_g \bLambda'_g + \bPsi_g\right)^{-1} \right\} + \sum_{g=1}^G n_g \ln \pi_g.
\end{split}\end{equation*}

The E-step on the first cycle of the $\left(k+1\right)$st iteration requires the calculation of 
$Q_1\left(\btheta_1; \btheta^{\left(k\right)}\right) = \mathbb{E}_{\btheta^{\left(k\right)}}\left[l_{c} \left(\btheta_1\right)|\mathcal{S}\right]$, which is the expected complete-data log-likelihood given the observed data and using the current estimate $\btheta^{\left(k\right)}$ for $\btheta$. 
In practice, it requires calculating
$\mathbb{E}_{\btheta^{\left(k\right)}}\left[Z_{ig}|\mathcal{S}\right]$; this step is achieved by replacing each $z_{ig}$ by $z_{ig}^{(k+1)}$, where
\begin{displaymath}
z_{ig}^{\left(k+1\right)} = \frac{\pi_j^{\left(k\right)}\phi\left(y_i|\bx_i; \mu\left(\boldsymbol{x}_i;\bbeta_g^{(k)}\right), \sigma_g^{2\left(k\right)}\right)   \phi\left(\bx_i|\bmu_g^{\left(k\right)},\bLambda_g^{\left(k\right)},\bPsi^{\left(k\right)}_g\right) }{\displaystyle\sum_{j=1}^G \pi_j^{\left(k\right)}\phi\left(y_i|\bx_i; \mu\left(\boldsymbol{x}_i;\bbeta_j^{(k)}\right), \sigma_j^{2\left(k\right)}\right)\phi \left(\bx_i|\bmu_j^{\left(k\right)},\bLambda_j^{\left(k\right)},\bPsi_j^{\left(k\right)}\right) }.
\end{displaymath}

For the M-step, the maximization of this complete-data log-likelihood yields
\begin{equation*}\begin{split}
\pi_g^{\left(k+1\right)} & = \frac{1}{n}\displaystyle\sum_{i=1}^n z_{ig}^{\left(k+1\right)} 
\\
\bmu_g^{\left(k+1\right)} & = \frac{1}{n_g} \sum_{i=1}^n z_{ig}^{\left(k+1\right)} \bx_i \nonumber
\\
\bbeta_{1g}^{\left(k+1\right)} & = \left[ \frac{1}{n_g} \sum_{i=1}^n z_{ig}^{\left(k+1\right)} y_i \left(\bx_i - \bmu_g^{\left(k+1\right)}\right)\right] \left[\frac{1}{n_g}\sum_{i=1}^n z_{ig}^{\left(k+1\right)} \bx'_i\bx_i - \bmu_g^{'\left(k+1\right)}\bmu_g^{\left(k+1\right)}\right]^{-1} 
\\
\beta_{0g}^{\left(k+1\right)} & = \frac{1}{n_g}\displaystyle\sum_{i=1}^n z_{ig}^{\left(k+1\right)} y_i-\bbeta_{1g}^{'\left(k+1\right)}\bmu_g^{\left(k+1\right)} 
\\
\sigma_g^{2\left(k+1\right)} & = \frac{1}{n_g} \sum_{i=1}^n z_{ig}^{\left(k+1\right)} \left\{ y_i-\left(\beta_{0g}^{\left(k+1\right)}+\bbeta_{1g}^{'\left(k+1\right)}\bx_i\right)\right\}^2,  
\end{split}\end{equation*}
where $n_g^{\left(k+1\right)}=\displaystyle\sum_{i=1}^n z_{ig}^{\left(k+1\right)}$. 
Following the notation in \citet{McLa:Peel:fini:2000}, we set $\btheta^{\left(k+1/2\right)}=\left\{\btheta_1^{\left(k+1\right)},\btheta_2^{\left(k\right)}\right\}$.

\subsection{Second cycle}\label{sec:second}

Here, $\btheta_2=\left\{\bSigma_g ; g=1, \ldots, G \right\}= \left\{\bLambda_g,\bPsi_g;g=1, \ldots, G\right\}$, where the missing data are the unobserved group labels $\boldsymbol{z}_i$ and the latent factors $\boldsymbol{u}_{ig}$, $i=1,\ldots,n$ and $g=1,\ldots,G$.
Therefore, the complete-data likelihood is
\begin{equation*}\begin{split}
L_{c2}(\btheta_2) &= \prod_{i=1}^n\prod_{g=1}^G\left[\phi\left(y_i|\bx_i, \bu_{ig}; \mu\left(\bx_i;\bbeta_g^{\left(k+1\right)}\right),\sigma_g^{2\left(k+1\right)} \right) \phi\left(\bx_i|\bu_{ig};\bmu_g^{\left(k+1\right)},\bSigma_g\right)\phi (\bu_{ig}) \pi_g^{\left(k+1\right)} \right]^{z_{ig}} \nonumber \\
&= \prod_{i=1}^n\prod_{g=1}^G\left[\phi\left(y_i|\bx_i; \mu\left(\bx_i;\bbeta_g^{\left(k+1\right)}\right),\sigma_g^{2\left(k+1\right)} \right) \phi\left(\bx_i|\bu_{ig};\bmu_g^{\left(k+1\right)},\bLambda_g,\bPsi_g\right) \phi (\bu_{ig}) \pi_g^{\left(k+1\right)} \right]^{z_{ig}},
\end{split}\end{equation*}
because $Y$ is conditionally independent of $\bU$ given $\bX=\bx$ and 
\begin{equation*}\begin{split}
\phi\left(\bx_i|\bu_{ig};\bmu_g^{\left(k+1\right)},\bPsi_g\right) &= \frac{1}{\left|2 \pi \bPsi_g\right|^{1/2}} \exp \left\{ - \frac{1}{2} \left(\bx_i - \bmu_g^{\left(k+1\right)}-\bLambda_g \bu_{ig}\right)' \bPsi_g^{-1} \left(\bx_i - \bmu_g^{\left(k+1\right)}- \bLambda_g \bu_{ig}\right) \right\}  \\ 
\phi \left(\bu_{ig}\right) &= \frac{1}{\left(2 \pi\right)^{q/2}} \exp \left\{ - \frac{1}{2}\bu_{ig}' \bu_{ig} \right\}. 
\end{split}\end{equation*}
Hence, the complete-data log-likelihood is
\begin{equation*}\begin{split}
l_{c2}\left(\btheta_2\right) &= - \frac{n\left(p+q+1\right)}{2} \ln \left(2\pi\right) - \frac{1}{2} \sum_{i=1}^n \sum_{g=1}^G z_{ig} \ln \sigma_g^{2\left(k+1\right)} + \nonumber \\
& -\frac{1}{2}\sum_{i=1}^n\sum_{g=1}^G z_{ig} \frac{\left(y_i-\beta_{0g}^{\left(k+1\right)}-\bbeta_{1g}^{'\left(k+1\right)}\bx_i\right)^2}{2 \hat{\sigma}_g^2}  + \sum_{g=1}^G n_g \ln \pi_g  +\frac{1}{2}\sum_{i=1}^n \sum_{g=1}^G z_{ig} \ln \left| \bPsi^{-1}_g\right| + \nonumber \\ 
& - \frac{1}{2} \sum_{i=1}^n\sum_{g=1}^G z_{ig}\tr\left\{ \left(\bx_i - \bmu_g^{\left(k+1\right)}- \bLambda_g \bu_{ig}\right) \left(\bx_i - \bmu_g^{\left(k+1\right)}- \bLambda_g \bu_{ig}\right)' \bPsi_g^{-1} \right\}, 
\label{L2(theta2)}
\end{split}\end{equation*}
where we set 
\begin{displaymath}
\bS_g^{\left(k+1\right)}=\frac{1}{n_g^{\left(k+1\right)}}\sum_{i=1}^n z_{ig}^{\left(k+1\right)}\left(\bx_i - \bmu_g^{\left(k+1\right)}\right) \left(\bx_i - \bmu_g^{\left(k+1\right)}\right)'. 	
\end{displaymath}

The E-step on the second cycle of the $\left(k+1\right)$st iteration requires the calculation of $Q_2\left(\btheta_2; \btheta^{(k+1/2)}\right) = \mathbb{E}_{\btheta^{(k+1/2)}} \left[l_{c2} \left(\btheta_2\right)| \mathcal{S}\right]$.
Therefore, we must calculate the following conditional expectations: $\mathbb{E}_{\btheta^{(k+1/2)}} \left( Z_{ig} | \mathcal{S}\right)$, $\mathbb{E}_{\btheta^{(k+1/2)}} \left( Z_{ig} \bU_{ig} | \mathcal{S}\right)$,
and $\mathbb{E}_{\btheta^{(k+1/2)}} \left( Z_{ig}\bU_{ig} \bU'_{ig} |\mathcal{S}\right)$. 
Based on \eqref{eq:U|xy}, these are given by
\begin{equation*}\begin{split}
\mathbb{E}_{\btheta^{(k+1/2)}} \left(Z_{ig} \bU_{ig} | \mathcal{S}\right) &=z_{ig}^{\left(k+1\right)}\bgamma_g^{\left(k\right)}\left(\bx_i- \bmu_g^{\left(k+1\right)}\right) \\
\mathbb{E}_{\btheta^{(k+1/2)}} \left(Z_{ig}\bU_{ig} \bU'_{ig} | \mathcal{S}\right)  &= z_{ig}^{\left(k+1\right)} \left\{\bI_q-\bgamma^{\left(k\right)}_g \bLambda^{\left(k\right)}_g+\bgamma^{\left(k\right)}_g \bS_g \bgamma^{'\left(k\right)}_g \right\} = z_{ig}^{\left(k+1\right)} \bTheta^{\left(k\right)}_g, 
\end{split}\end{equation*}
where 
\begin{eqnarray}
\bgamma^{\left(k\right)}_g &=& \bLambda'^{\left(k\right)}_g\left(\bLambda^{\left(k\right)}_g\bLambda'^{\left(k\right)}_g+\bPsi^{\left(k\right)}_g\right)^{-1}\\ \label{gamma^k}
\bTheta^{\left(k\right)}_g &=& 
\bI_q-\bgamma^{\left(k\right)}_g \bLambda^{\left(k\right)}_g+ \bgamma^{\left(k\right)}_g \bS_g^{\left(k+1\right)}
\bgamma'^{\left(k\right)}_g. 
\label{Theta^k}
\end{eqnarray}
Thus, the $g$th term of the expected complete-data log-likelihood $Q_2\left(\btheta_2; \btheta^{(k+1/2)}\right)$  becomes 
\begin{equation}\begin{split}
Q_2&\left(\bLambda_g, \bPsi_g; \btheta^{(k+1/2)}\right) = \mbox{ C}(\btheta_1^{\left(k+1\right)}) +\frac{1}{2} n_g^{\left(k+1\right)}\ln | \bPsi^{-1}_g| - \frac{1}{2} n_g^{\left(k+1\right)} \tr\left\{ \bS_g^{\left(k+1\right)} \bPsi_g^{-1} \right\}\\ 
& + 
n_g^{\left(k+1\right)}\text{tr}\left\{\bLambda_g \bgamma_g^{\left(k\right)}\bS_g^{\left(k+1\right)} \bPsi^{-1}_g \right\} 
 -\frac{1}{2} n_g^{\left(k+1\right)}\text{tr}\left\{\bLambda_g'\bPsi_g^{-1} \bLambda_g \bTheta^{\left(k\right)}_g\right\} ,
\label{eq:expected complete-data log-likelihood2}
\end{split}\end{equation}
where $\text{C}\left(\btheta_1^{\left(k+1\right)}\right)$ denotes the terms in \eqref{L2(theta2)} that do not depend on $\btheta_2$. 
Then \eqref{eq:expected complete-data log-likelihood2} is maximized for $\left\{\hat{\bLambda}, \hat{\bPsi}\right\}$, satisfying
\begin{equation*}\begin{split}
\frac{\partial Q_2}{\partial \bLambda_g} &= n_g^{\left(k+1\right)} \bPsi_g^{-1} \bS_g^{\left(k+1\right)} \bgamma_g^{'\left(k\right)} - n_g^{\left(k+1\right)} \bPsi_g^{-1} \bLambda_g  \bTheta^{\left(k\right)}_g = \bzero 
\\
\frac{\partial Q_2}{\partial \bPsi^{-1}_g} &= \frac{1}{2}n_g^{\left(k+1\right)}\bPsi_g-\frac{1}{2}n_g^{\left(k+1\right)}\bS_g^{\left(k+1\right)} + n_g^{\left(k+1\right)} \bS_g^{'\left(k+1\right)} \bgamma_g^{'\left(k\right)} \bLambda_g'-\frac{1}{2} n_g^{\left(k+1\right)}\bLambda_g\bTheta_g^{\left(k\right)}\bLambda_g' = \bzero.
\end{split}\end{equation*}
Therefore,
\begin{eqnarray}
 \bS_g^{\left(k+1\right)} \bgamma_g^{'\left(k\right)} - \bLambda_g  \bTheta^{\left(k\right)}_g  & = &\bzero 
 \label{Q2Lambda} 
 \\
\bPsi_g-\bS_g^{\left(k+1\right)} +2\bS_g^{'\left(k+1\right)} \bgamma_g^{'\left(k\right)} \bLambda_g'-\bLambda_g\bTheta_g^{\left(k\right)}\bLambda_g'  & = & \bzero . 
\label{Q2Psi}
\end{eqnarray}
From \eqref{Q2Lambda}, we get
\begin{equation}
\hat{\bLambda}_g  = \bS^{\left(k+1\right)}_g \bgamma'^{\left(k\right)}_g\bTheta_g^{-1}, \label{update_Lambdag}
\end{equation}
and substituting in \eqref{Q2Psi} we get
\begin{equation*}\begin{split}
\bPsi_g -\bS_g^{\left(k+1\right)}+2 \bS_g^{\left(k+1\right)} \bgamma'^{\left(k\right)}_g \left(\bS_g^{\left(k+1\right)}\bgamma'^{\left(k\right)}_g\bTheta_g^{-1}\right)'-\left(\mathbf{S}_g\hat{\bgamma}'_g\bTheta_g^{-1}\right)\bTheta_g\left(\mathbf{S}_g\hat{\bgamma}'_g\bTheta_g^{-1}\right)' = \bzero
\end{split}\end{equation*}
which yields
\begin{equation}
 \hat{\bPsi}_g  =\text{diag}\left\{ \bS^{\left(k+1\right)}_g- \hat{\bLambda}_g \hat{\bgamma}_g\bS^{\left(k+1\right)}_g\right\}. \label{update_Psig}
\end{equation}

Hence, the maximum likelihood estimates for $\bLambda$ and $\bPsi$ are obtained by iteratively computing
\begin{equation*}\begin{split}
\bLambda_g^{+} &= \bS^{\left(k+1\right)}_g \bgamma_g^{'} \bTheta_g^{-1}  
\\ 
\bPsi^{+}_g  &= \text{diag}\left\{ \bS^{\left(k+1\right)}_g- \bLambda_g^{+} \bgamma_g\bS^{\left(k+1\right)}_g\right\}, 
\end{split}\end{equation*}
where the superscript $^{+}$  denotes the update estimate. 
Using \eqref{gamma^k} and \eqref{Theta^k}, we get
\begin{eqnarray}
\bgamma_g^+ & = & \bLambda'^{+}_g  \left(\bLambda^+_g \bLambda'^{+}_g +\bPsi^+_g\right)^{-1} 
\nonumber 
\\
\bTheta^+_g & = & \bI_q-\bgamma^+_g \bLambda^+_g+ \bgamma^+_g \bS_g^{\left(k+1\right)} \bgamma'^{+}_g . \label{Theta^+_g} 
\end{eqnarray}


\subsection{Outline of the algorithm}
\label{subsec:CovarianceX_update}

In summary, the procedure can be described as follows. For a given initial guess $\btheta^{\left(0\right)}$, on the $\left(k+1\right)$st iteration, the
algorithm carries out the following steps for $g=1,\ldots,G$:
\begin{enumerate}
\item Compute $\pi^{\left(k+1\right)}_g, \bmu^{\left(k+1\right)}_g, \bbeta^{\left(k+1\right)}_g, \sigma^{2\left(k+1\right)}_g$;
\item Set $\bLambda_g \leftarrow \bLambda^{\left(k\right)}_g$ and $\bPsi \leftarrow \bPsi^{\left(k\right)}_g$, and compute $\bgamma_g$ and $\bTheta_g$;
\item Repeat the following steps  until convergence on $\bLambda_g$ and $\bPsi_g$:
\begin{enumerate}
\item Set $\bLambda_g^{+} \leftarrow  \bS^{\left(k+1\right)}_g \bgamma'_g \bTheta_g^{-1}$ and $\bPsi^{+}_g  \leftarrow \text{diag}\left\{ \bS^{\left(k+1\right)}_g - \bLambda_g^{+} \bgamma_g\bS^{\left(k+1\right)}_g\right\}$;
\item Set $\bgamma^{+}_g \leftarrow\bLambda'^{+}_g \left(\bLambda_g^{+} \bLambda'^{+}_g +\bPsi^{+}_g\right)^{-1}$ and $\bTheta^{+}_g \leftarrow \bI_q-\bgamma^{+}_g \bLambda^{+}_g+ \bgamma^{+}_g \bS_g^{\left(k+1\right)} \bgamma'^{+}_g$;
\item Set $\bLambda_g \leftarrow \bLambda_g^{+}$, $\bPsi_g \leftarrow \bPsi_g^+$, $\bgamma_g \leftarrow \bgamma^{+}_g$, and $\bTheta_g \leftarrow \bTheta^{+}_g$.
\end{enumerate}
\end{enumerate}


\subsection{AECM initialization: a 5-step procedure}
\label{subsec:AECM initialization}

The choice of starting values is a well known and important issue with respect to EM-based algorithms.
The standard approach consists of selecting a value for $\boldsymbol{\theta}^{\left(0\right)}$.
An alternative method, more natural in the authors' opinion, consists of choosing a value for $\boldsymbol{z}_i^{\left(0\right)}$, $i=1,\ldots,n$ \citep[see][p.~54]{McLa:Peel:fini:2000}.   
Within this approach, and due to the hierarchical structure of the CWFA family of parsimonious models, we propose a 5-step hierarchical initialization procedure. 

\begin{figure}[!ht]
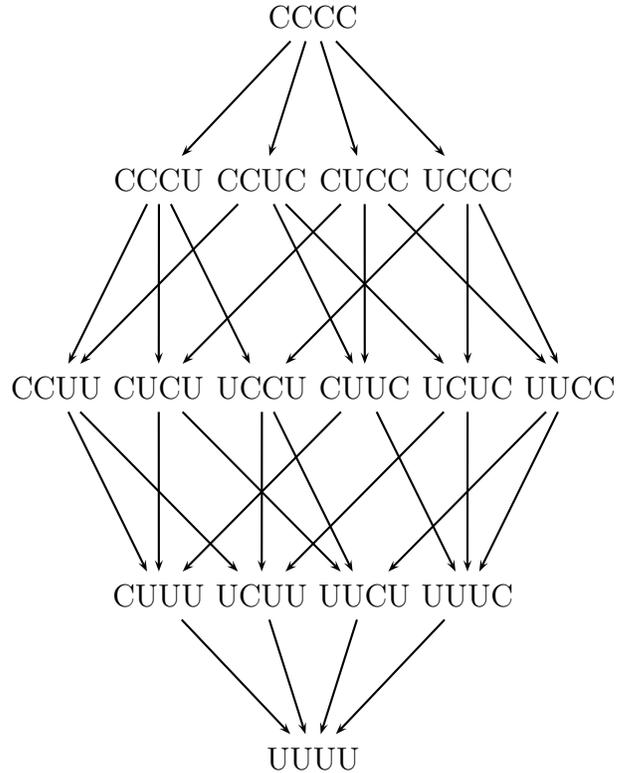

\centering
$
\begin{array}{cccccc}
\multicolumn{6}{c}{\rnode{CCCC}{\text{CCCC}}} \\[1.4cm]
& \multicolumn{1}{c}{\rnode{CCCU}{\text{CCCU}}} & \multicolumn{1}{c}{\rnode{CCUC}{\text{CCUC}}} & \multicolumn{1}{c}{\rnode{CUCC}{\text{CUCC}}} & \multicolumn{1}{c}{\rnode{UCCC}{\text{UCCC}}} & \\[2cm]
\multicolumn{1}{c}{\rnode{CCUU}{\text{CCUU}}} & \multicolumn{1}{c}{\rnode{CUCU}{\text{CUCU}}} & \multicolumn{1}{c}{\rnode{UCCU}{\text{UCCU}}} & \multicolumn{1}{c}{\rnode{CUUC}{\text{CUUC}}} & \multicolumn{1}{c}{\rnode{UCUC}{\text{UCUC}}} & \multicolumn{1}{c}{\rnode{UUCC}{\text{UUCC}}} \\[2cm]  & \multicolumn{1}{c}{\rnode{CUUU}{\text{CUUU}}} & \multicolumn{1}{c}{\rnode{UCUU}{\text{UCUU}}} & \multicolumn{1}{c}{\rnode{UUCU}{\text{UUCU}}} & \multicolumn{1}{c}{\rnode{UUUC}{\text{UUUC}}} &\\[1.4cm]
\multicolumn{6}{c}{\rnode{UUUU}{\text{UUUU}}}\\
\end{array}
$
\psset{nodesep=5pt}
\ncline{->}{CCCC}{CCCU}
\ncline{->}{CCCC}{CCUC}
\ncline{->}{CCCC}{CUCC}
\ncline{->}{CCCC}{UCCC}
\ncline{->}{CCCU}{CCUU}
\ncline{->}{CCCU}{CUCU}
\ncline{->}{CCCU}{UCCU}
\ncline{->}{CCUC}{UCUC}
\ncline{->}{CCUC}{CUUC}
\ncline{->}{CCUC}{CCUU}
\ncline{->}{CUCC}{UUCC}
\ncline{->}{CUCC}{CUUC}
\ncline{->}{CUCC}{CUCU}
\ncline{->}{UCCC}{UUCC}
\ncline{->}{UCCC}{UCUC}
\ncline{->}{UCCC}{UCCU}
\ncline{->}{CCUU}{UCUU}
\ncline{->}{CCUU}{CUUU}
\ncline{->}{CUCU}{UUCU}
\ncline{->}{CUCU}{CUUU}
\ncline{->}{CUUC}{UUUC}
\ncline{->}{CUUC}{CUUU}
\ncline{->}{UUCC}{UUUC}
\ncline{->}{UUCC}{UUCU}
\ncline{->}{UCUC}{UUUC}
\ncline{->}{UCUC}{UCUU}
\ncline{->}{UCCU}{UCUU}
\ncline{->}{UCCU}{UUCU}
\ncline{->}{CUUU}{UUUU}
\ncline{->}{UCUU}{UUUU}
\ncline{->}{UUCU}{UUUU}
\ncline{->}{UUUC}{UUUU}
\caption{
Relationships among the models in the 5-step hierarchical initialization procedure.
Arrows are oriented from the model used to initialize to the model to be estimated.   
}
\label{fig:hierarchy}
\end{figure}

For a fixed number of groups $G$ , let $\boldsymbol{z}_i^{\left(0\right)}$, $i=1,\ldots,n$, be the initial classification for the AECM algorithm, so that $z_{ig}^{\left(0\right)}\in\left\{0,1\right\}$ and $\sum_gz_{ig}^{\left(0\right)}=1$.
The set $\left\{\bz_i^{\left(0\right)};i=1, \ldots, n\right\}$ can be obtained either through some clustering procedure (here we consider the $k$-means method) or by random initialization, for example by sampling from a multinomial distribution with probabilities $\left(1/G,\ldots,1/G\right)$.
Then, at the first step of the procedure, the most constrained CCCC model is estimated from these starting values.
At the second step, the resulting (AECM-estimated) $\hat{z}_{ig}$ are taken as the starting group membership labels to initialize the AECM-algorithm of the four models $\left\{\text{UCCC}, \text{CUCC}, \text{CCUC}, \text{CCCU}\right\}$ obtained by relaxing one of the four constraints.
At the third step, the AECM-algorithm for each of the six models $\left\{\text{CCUU}, \text{CUCU}, \text{UCCU}, \text{CUUC}, \text{UCUC}, \text{UUCC}\right\}$ with two constraints is initialized using the $\hat{z}_{ig}$ from the previous step and the model with the highest likelihood. 
For example, to initialize CCUU we use the $\hat{z}_{ig}$ from the model having the highest likelihood between CCCU and CCUC. 
In this fashion, the initialization procedure continues according to the scheme displayed in \figurename~\ref{fig:hierarchy}, until the least constrained model UUUU is estimated at the fifth step.
 
For all of the models in the CWFA family, in analogy with \cite{McNi:Murp:Pars:2008}, the initial values for the elements of $\boldsymbol{\Lambda}_g$ and $\boldsymbol{\Psi}_g$ are generated from the eigen-decomposition of $\boldsymbol{S}_g$ as follows.
The $\boldsymbol{S}_g$ are computed based on the values of $z_{ig}^{\left(0\right)}$.
The eigen-decomposition of each $\boldsymbol{S}_g$ is obtained using the Householder reduction and the QL method (details given by \citealp{Pres:Teuk:Vett:Flan:Nume:1992}). 
Then the initial values of the elements of $\boldsymbol{\Lambda}_g$ are set as $\lambda_{ij}=\sqrt{d_j}\rho_{ij}$, where $d_j$ is the $j$th largest eigenvalue of $\boldsymbol{S}_g$ and $\rho_{ij}$ is the $i$th element of the eigenvector corresponding to the $j$th largest eigenvalue of $\boldsymbol{S}_g$, where $i\in\left\{1,2,\ldots,d\right\}$ and $j\in\left\{1,2,\ldots,q\right\}$.
The $\boldsymbol{\Psi}_g$ are then initialized as $\boldsymbol{\Psi}_g=\diag\left(\boldsymbol{S}_g-\boldsymbol{\Lambda}_g\boldsymbol{\Lambda}_g'\right)$.

\subsection{Convergence criterion}
\label{subsec:Convergence criterion}

The Aitken acceleration procedure \citep{Aitk:OnBe:1926} is used to estimate the asymptotic maximum of the log-likelihood at each iteration of the AECM algorithm. 
Based on this estimate, a decision is made about whether the algorithm has reached convergence, i.e., whether the log-likelihood is sufficiently close to its estimated asymptotic value. 
The Aitken acceleration at iteration $k$ is given by
\begin{displaymath}
	a^{\left(k\right)}=\frac{l^{\left(k+1\right)}-l^{\left(k\right)}}{l^{\left(k\right)}-l^{\left(k-1\right)}},
\end{displaymath}
where $l^{\left(k+1\right)}$, $l^{\left(k\right)}$, and $l^{\left(k-1\right)}$ are the (observed-data) log-likelihood values from iterations $k+1$, $k$, and $k-1$, respectively. 
Then, the asymptotic estimate of the log-likelihood at iteration $k + 1$ is
\begin{displaymath}	l_{\infty}^{\left(k+1\right)}=l^{\left(k\right)}+\frac{1}{1-a^{\left(k\right)}}\left(l^{\left(k+1\right)}-l^{\left(k\right)}\right)
\end{displaymath}
\citep{Bohn:Diet:Scha:Schl:Lind:TheD:1994}.
In the analyses in Section \ref{sec:Data analyses}, we stop our algorithms when $l_{\infty}^{\left(k+1\right)}-l^{\left(k\right)}<\epsilon$ (\citealp{Bohn:Diet:Scha:Schl:Lind:TheD:1994,McNi:Murp:McDa:Fros:Seri:2010}). Note that we use $\epsilon=0.05$ for the analyses herein.

\section{Model selection and performance assessment}\label{sec:Model selection and performance assessment}

\subsection{Model selection}
\label{subsec:Model selection}

The CWFA model, in addition to $\btheta$, is also characterized by the number of latent factors $q$ and by the number of mixture components $g$. So far, these quantities have been treated as \textit{a priori} fixed. 
Nevertheless, the estimation of these is required, for practical purposes, when choosing a relevant model.

For model-based clustering and classification, several model selection criteria are used, such as the Bayesian information criterion \citep[BIC;][]{Schw:Esti:1978}, the integrated completed likelihood \citep[ICL;][]{Bier:Cele:Gova:Asse:2000}, and the Akaike information criterion \citep[AIC;][]{Saka:Ishi:Kita:Akai:1983}. 
Among these, the BIC is the most predominant in the literature and is given by 
$$\text{BIC}=2l\left(\hat{\btheta}\right) - \eta \ln\left(n\right),$$
where $l\left(\hat{\btheta}\right)$ is the (maximized) observed-data log-likelihood and $\eta$ is the number of free parameters. This is the model selection criterion used in the analyses of Section~\ref{sec:Data analyses}.

\subsection{Adjusted Rand index}
\label{subsec:Adjusted Rand index}

Although the data analyses of Section~\ref{sec:Data analyses} are mainly conducted as clustering examples, the true classifications are actually known for these data. 
In these examples, the adjusted Rand index \citep[ARI;][]{Hube:Arab:Comp:1985} is used to measure class agreement. The original Rand Index \citep[RI;][]{Rand:Obje:1971} is based on pairwise comparisons and is obtained by dividing the number of pair agreements (observations that should be in the same group and are, plus those that should not be in the same group and are not) by the total number of pairs. 
The ARI corrects the RI to account for agreement by chance: a value of `1' indicates perfect agreement, `0' indicates random classification, and negative values indicate a classification that is worse than would be expected by guessing.

\section{Data analyses}
\label{sec:Data analyses}

This section presents the application of the family of parsimonious linear Gaussian models to both artificial and real data sets. Code for the AECM algorithm, described in this paper, was written in the {\sf R} computing environment \citep{R}. 

\subsection{Simulated data}
\label{subsec:Simulated data}


\subsubsection{Example 1}
\label{Ex:ExampleV2}

\begin{figure}[!ht]
\begin{center}
\resizebox{0.8\textwidth}{!}{\includegraphics{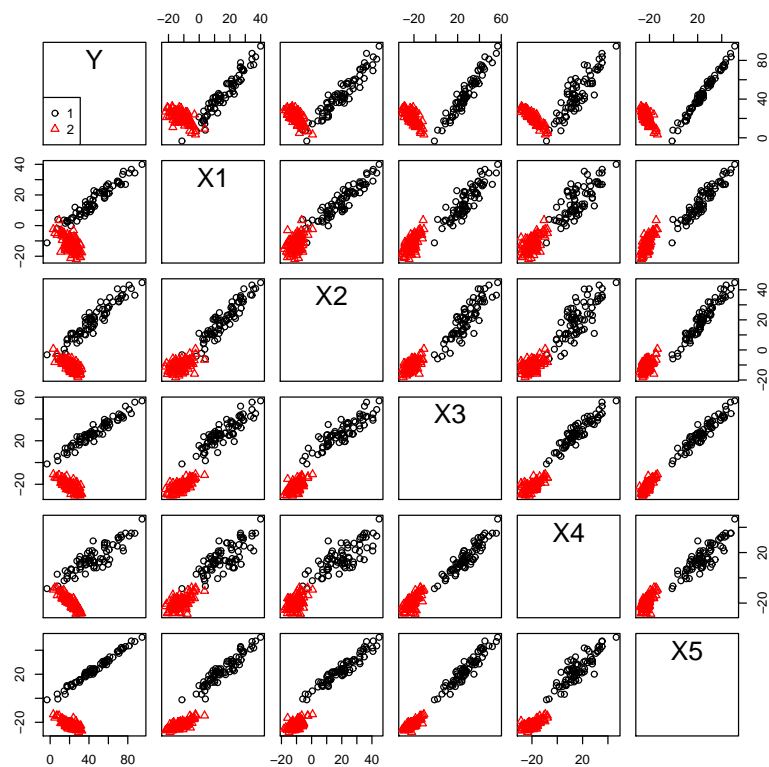}}
\caption{Example~\ref{Ex:ExampleV2}: scatterplot matrix of the simulated data.}
\label{fig:scatterdata_V2}
\end{center}
\end{figure}

The first data set consists of a sample of size $n=175$ drawn from model UUCU with $G=2$, $n_1=75$, $n_2=100$, $d=5$, and $q=2$ (see \figurename~\ref{fig:scatterdata_V2} for details). 
The parameters used for the simulation of the data are given in \tablename~\ref{tab:parameters of Example 1} (see \appendixname~\ref{app:Example 1} for details on the covariance matrices $\bSigma_g$, $g=1,\ldots,G$).
\begin{table}[!ht]
\caption{
True and estimated parameters for the simulated data of Example~1.
\label{tab:parameters of Example 1}
}
\centering
\subtable[Means of $\bX$]{
\begin{tabular}{ccrrrrrcrrrrr}
\toprule
      && \multicolumn{5}{c}{$\boldsymbol{\mu}_g$} & & \multicolumn{5}{c}{$\hat{\boldsymbol{\mu}}_g$} \\
Group && \multicolumn{1}{c}{$X_1$} & \multicolumn{1}{c}{$X_2$} & \multicolumn{1}{c}{$X_3$} & \multicolumn{1}{c}{$X_4$} & \multicolumn{1}{c}{$X_5$}  & & \multicolumn{1}{c}{$X_1$} & \multicolumn{1}{c}{$X_2$} & \multicolumn{1}{c}{$X_3$} & \multicolumn{1}{c}{$X_4$} & \multicolumn{1}{c}{$X_5$} \\
\midrule
1 &&  14.00 &  18.00 &  25.00 &  14.00 &  22.00 & &  15.88 &  19.94 &  27.48 &  15.81 &  23.93\\
2 && -12.00 & -10.00 & -22.00 & -20.00 & -22.00 & & -11.95 & -10.36 & -22.00 & -19.67 & -22.03\\
\bottomrule
\end{tabular}
}
\subtable[Slopes]{
\begin{tabular}{ccrrrrrcrrrrr}
\toprule
      && \multicolumn{5}{c}{$\bbeta_{1g}$}     & & \multicolumn{5}{c}{$\hat{\bbeta}_{1g}$} \\
Group && \multicolumn{1}{c}{$X_1$} & \multicolumn{1}{c}{$X_2$} & \multicolumn{1}{c}{$X_3$} & \multicolumn{1}{c}{$X_4$} & \multicolumn{1}{c}{$X_5$}  & & \multicolumn{1}{c}{$X_1$} & \multicolumn{1}{c}{$X_2$} & \multicolumn{1}{c}{$X_3$} & \multicolumn{1}{c}{$X_4$} & \multicolumn{1}{c}{$X_5$} \\
\midrule
1 &&  0.47 &   0.02 &   0.42 &   0.03 &   0.87 & &  0.50 &  0.03 &   0.46 &   0.02 &  0.81 \\
2 && -0.02 &  -0.63 &  -0.05 &  -0.85 &  -0.03 & & -0.04 & -0.57 &  -0.01 &  -0.85 & -0.18 \\
\bottomrule
\end{tabular}
}
\subtable[Conditional std. deviations]{
\begin{tabular}{ccrcr}
\toprule
Group && \multicolumn{1}{c}{$\sigma_g$} && \multicolumn{1}{c}{$\hat{\sigma}_g$} \\
\midrule
1 && 2.00 && 1.24\\
2 && 4.00 && 3.79\\
\bottomrule
\end{tabular}
}
\quad
\subtable[Intercepts]{
\begin{tabular}{ccrcr}
\toprule
Group && \multicolumn{1}{c}{$\beta_{0g}$} && \multicolumn{1}{c}{$\hat{\beta}_{0g}$} \\
\midrule
1 &&  4.50 &&  4.34\\
2 && -4.20 && -6.35\\
\bottomrule
\end{tabular}
}
\end{table}

All of the sixteen CWFA models were fitted to the data for $G\in\{2,3\}$ and $q\in\{1,2\}$, resulting in a total of 64 models. As noted above (Section~\ref{subsec:AECM initialization}), initialization of the $\bz_i$, $i=1,\ldots,n$, for the most constrained model (CCCC), and for each combination $\left(G,q\right)$, was done using the $k$-means algorithm according to the \texttt{kmeans} function of the {\sf R} package \texttt{stats}. 
The remaining 15 models, for each combination $\left(G,q\right)$, were initialized using the 5-step hierarchical initialization procedure described in Section~\ref{subsec:AECM initialization}. 
The BIC values for all 64 models were computed and the model with the largest BIC value was selected as the best model. In this example, the model corresponding to the largest BIC value (-5845.997) was a two component ($G=2$) UUCU model with two latent factors ($q=2$), the same as the model used to generate the data. 
The selected model gave a perfect classification and the estimated parameters were very close to the parameters used for data simulation (see \tablename~\ref{tab:parameters of Example 1} and \appendixname~\ref{app:Example 1}).

\begin{figure}[!ht]
\begin{center}
\resizebox{0.6\textwidth}{!}{\includegraphics{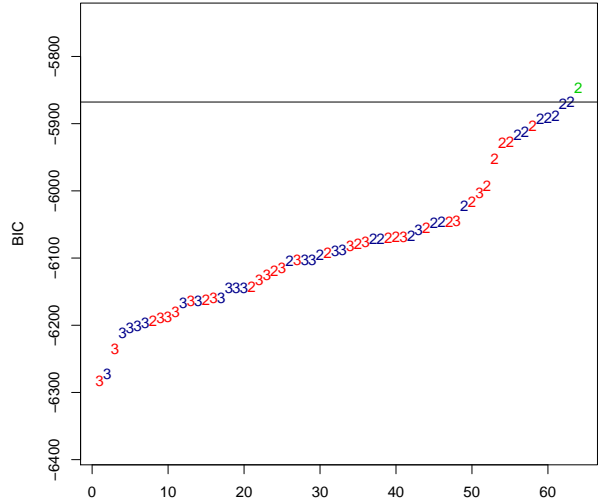}} 
\caption{Group-factor plot of BIC values sorted in an increasing order for Example~1. 
Numbers denote the selected number $G$ of groups and colours denote the number $q$ of latent factors (red: $q=1$, blue: $q=2$). 
The green number indicates the true model.}
\label{fig:BICsortqg_V}
\end{center}
\end{figure}

\figurename~\ref{fig:BICsortqg_V} shows the BIC values of all 64 models sorted in an increasing order, where numbers denote the selected number $G$ of groups and colours denote the number $q$ of latent factors.
The horizontal line separates the models with a BIC value within 1\% of the maximum (over the 64 models) BIC value (hereafter simply referred to as the `1\% line').
This graphical representation will be referred to as the `group-factor plot' of BIC values.  
Here, as mentioned earlier, the model with the largest BIC was UUCU (with $G=2$ and $q=2$).
The subsequent two models, those above the 1\% line, were UUUU with $G=2$ and $q=2$ (BIC equal to $-5867.006$) and CUCU with $G=2$ and $q=2$ (BIC equal to $-5869.839$). 
These two models are structurally very close to the true UUCU model and also yielded perfect classification. 
It should also be noted that most of the models with high BIC values have $G=2$ and $q=2$.

\subsubsection{Example 2}
\label{Ex:ExampleV3}

\begin{figure}[!ht]
\begin{center}
\resizebox{0.8\textwidth}{!}{\includegraphics{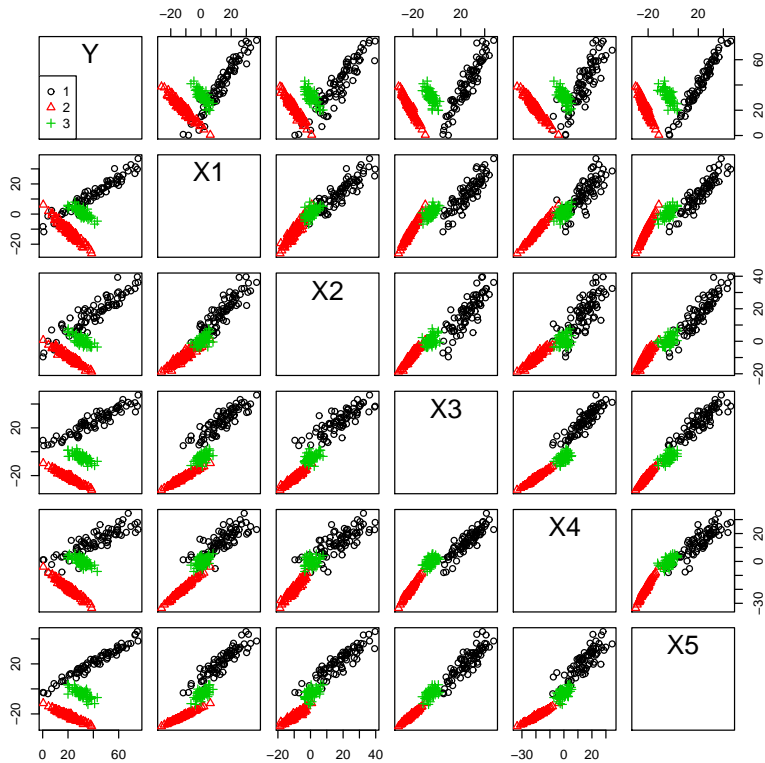}}
\caption{Scatterplot matrix of the simulated data for Example~2.}
\label{fig:scatterdata_V3}
\end{center}
\end{figure}

For the second data set, a sample of size $n=235$ was drawn from the CUUC model with $G=3$ groups (of size $n_1=75$, $n_2=100$, and $n_3=60$) and $q=2$ latent factors (see \figurename~\ref{fig:scatterdata_V3}). 

All 16 CWFA models were fitted to the data for $G\in\{2,3,4\}$ and $q\in\{1,2\}$, resulting in 96 different models. 
The algorithm was initialized in the same way as for Example~2. 
The model with the highest BIC (-6579.116) was CUUC with $G=3$ and $q=2$, resulting in a perfect classification. 
The estimated parameters of this model were very close to the true ones (\tablename~\ref{tab:parameters of Example 2} and \appendixname~\ref{appendixe2}).
\begin{table}[!ht]
\caption{
True and estimated parameters for the simulated data of Example~2.
\label{tab:parameters of Example 2}
}
\centering
\subtable[Means of $\bX$]{
\begin{tabular}{ccrrrrrcrrrrr}
\toprule
      && \multicolumn{5}{c}{$\boldsymbol{\mu}_g$} & & \multicolumn{5}{c}{$\hat{\boldsymbol{\mu}}_g$} \\
Group && \multicolumn{1}{c}{$X_1$} & \multicolumn{1}{c}{$X_2$} & \multicolumn{1}{c}{$X_3$} & \multicolumn{1}{c}{$X_4$} & \multicolumn{1}{c}{$X_5$}  & & \multicolumn{1}{c}{$X_1$} & \multicolumn{1}{c}{$X_2$} & \multicolumn{1}{c}{$X_3$} & \multicolumn{1}{c}{$X_4$} & \multicolumn{1}{c}{$X_5$} \\
\midrule
1&&   0.00 &   0.00 &  -5.00 &   0.00 &  -4.00 & &  0.82 &   0.48 &  -5.09 &  -0.21 &  -3.75 \\
2&&  14.00 &  18.00 &  25.00 &  14.00 &  22.00 & & 13.64 &  17.44 &  25.44 &  14.25 &  21.44 \\
3&& -12.00 & -10.00 & -22.00 & -20.00 & -22.00 & &-12.33 & -10.22 & -22.25 & -20.24 & -22.21 \\
\bottomrule
\end{tabular}
}
\subtable[Slopes]{
\begin{tabular}{ccrrrrrcrrrrr}
\toprule
      && \multicolumn{5}{c}{$\bbeta_{1g}$}     & & \multicolumn{5}{c}{$\hat{\bbeta}_{1g}$} \\
Group && \multicolumn{1}{c}{$X_1$} & \multicolumn{1}{c}{$X_2$} & \multicolumn{1}{c}{$X_3$} & \multicolumn{1}{c}{$X_4$} & \multicolumn{1}{c}{$X_5$}  & & \multicolumn{1}{c}{$X_1$} & \multicolumn{1}{c}{$X_2$} & \multicolumn{1}{c}{$X_3$} & \multicolumn{1}{c}{$X_4$} & \multicolumn{1}{c}{$X_5$} \\
\midrule
1&&  -0.41 & -0.87 & -0.22 & -0.62 & -0.06&& -0.34 & -0.82 & -0.32 & -0.66 & -0.09 \\
2&&   0.47 &  0.02 &  0.42 &  0.03 &  0.87&&  0.51 &  0.00 &  0.38 &  0.05 &  0.84 \\
3&&  -0.02 & -0.63 & -0.05 & -0.85 & -0.03&& -0.04 & -0.68 & -0.36 & -0.44 & -0.18 \\
\bottomrule
\end{tabular}
}
\subtable[Conditional std. deviations]{
\begin{tabular}{ccrcr}
\toprule
Group && \multicolumn{1}{c}{$\sigma_g$} && \multicolumn{1}{c}{$\hat{\sigma}_g$} \\
\midrule
1 && 2.00 && 2.30 \\
2 && 2.00 && 2.30 \\
3 && 2.00 && 2.30 \\
\bottomrule
\end{tabular}
}
\quad
\subtable[Intercepts]{
\begin{tabular}{ccrcr}
\toprule
Group && \multicolumn{1}{c}{$\beta_{0g}$} && \multicolumn{1}{c}{$\hat{\beta}_{0g}$} \\
\midrule
1&& 30.00 && 29.39\\
2&&  4.50 &&  5.31\\
3&& -4.20 && -6.69\\
\bottomrule
\end{tabular}
}
\end{table}

\begin{figure}[!ht]
\begin{center}
\includegraphics[width=0.7\textwidth]{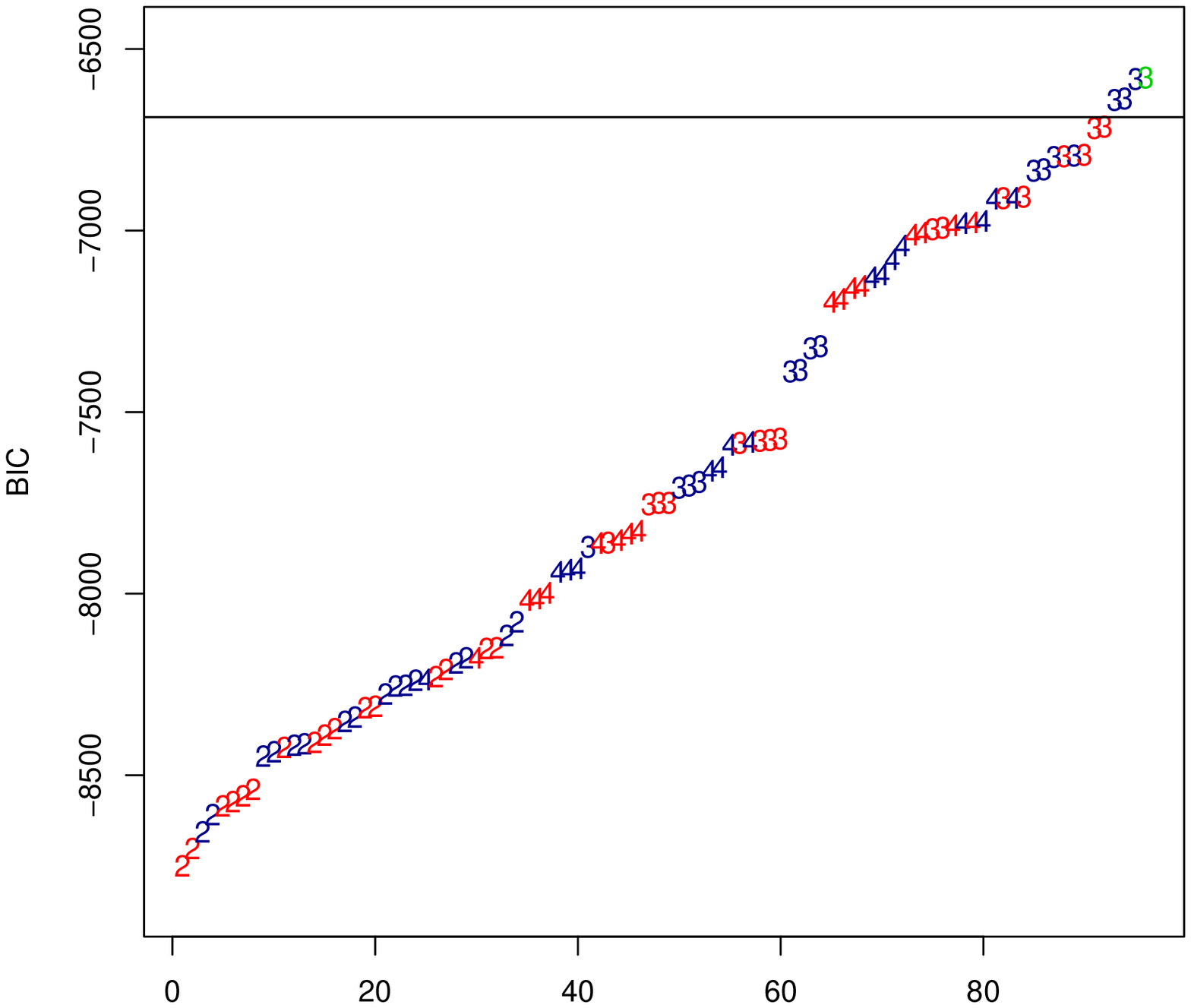} 
\caption{Group-factor plot of BIC values sorted in increasing order for Example~2. 
Numbers denote the selected number $G$ of groups and colours denote the number $q$ of latent factors (red: $q=1$, blue: $q=2$). 
The green number indicates the true model.}
\label{fig:BICsortqg_V3}
\end{center}
\end{figure}

\figurename~\ref{fig:BICsortqg_V3} shows the group-factor plot of BIC values for all 96 models. 
The other three models above the 1\% line are UUUC (BIC$=-6583.692$), CUUU (BIC$=-6637.222$), and UUUU (BIC$=-6641.798$), all with $G=3$ and $q=2$.
Thus, these models are congruent, with respect to the true one, in terms of $G$ and $q$. 
Moreover, they had a covariance structure more similar to the true one (CUUC) and also yielded perfect classification.

\subsection{The f.voles data set}
\label{subsec:real data}

In addition to the simulated data analyses of Section~\ref{subsec:Simulated data}, the family of CWFAs was also applied to a real data set for both clustering and classification purposes.

The \texttt{f.voles} data set, detailed in \citet[][Table~5.3.7]{Flur:Afir:1997} and available in the \texttt{Flury} package for {\sf R}, consists of measurements of female voles from two species, \textit{M.~californicus} and \textit{M.~ochrogaster}. The data consist of 86 observations for which we have a binary variable $\mathsf{Species}$ denoting the species ($n_1=45$ \textit{Microtus~ochrogaster} and $n_1=41$ \textit{M.~californicus}), a variable $\mathsf{Age}$ measured in days, and  six remaining variables related to skull measurements. The names of the variables are the same as in the original analysis of this data set by \citet{Airo:Hoff:Acom:1984}: $\mathsf{L}_2=\text{condylo-incisive length}$, $\mathsf{L}_9=\text{length of incisive foramen}$, $\mathsf{L}_7=\text{alveolar length of upper molar tooth row}$, $\mathsf{B}_3=\text{zygomatic width}$, $\mathsf{B}_4=\text{interorbital width}$, and $\mathsf{H}_1=\text{skull height}$. 
All of the variables related to the skull are measured in units of 0.1 mm.

The purpose of \citet{Airo:Hoff:Acom:1984} was to study age variation in \textit{M.~californicus} and \textit{M.~ochrogaster} and to predict age on the basis of the skull measurements. 
For our purpose, we assume the data are unlabelled with respect to $\mathsf{Species}$ and that our are interest is in evaluating clustering and classification using the family of CWFA models as well as comparing the algorithm with some well-established mixture model-based techniques. 
Therefore, $\mathsf{Age}$ can be considered the natural $Y$ variable and the $d=6$ skull measurements can be considered as the $\bX$ variable for the CWFA framework.

\subsubsection{Clustering}
\label{subsubsec:Clustering on real data}

All sixteen linear Gaussian CWFA models were fitted --- assuming no known group membership --- for $G\in\{2,\ldots,5\}$ components and $q\in\{1,2,3\}$ latent factors, resulting in total of 192 different models. 
The model with the largest BIC value was CCCU with $G=3$ and $q=1$, with a BIC of $-3837.698$ and an ARI of 0.72. \tablename~\ref{tab:CWFA} displays the clustering results from this model.
\begin{table}[!ht]
\caption{
Clustering of \texttt{f.voles} data using three different clustering approaches.
}\label{tab:f.voles clustering}
\centering
\subtable[CWFA]{\label{tab:CWFA}
\begin{tabular}{cccc}
\toprule
\backslashbox{TRUE}{Est.}     	&	 1	&	 2	&	 3 \\
\midrule															
\textit{ochrogaster}  & 24	&	21	&	 --\\
\textit{californicus} & -- & --	&	41 \\
\bottomrule	
\end{tabular}
}
\quad
\subtable[PGMM]{
\label{tab:PGMM}
\begin{tabular}{cccc}
\toprule
\backslashbox{TRUE}{Est.}     	&	 1	&	 2	&	 3 \\
\midrule															
\textit{ochrogaster}  & 34	&	 9	&	 \textbf{2} \\
\textit{californicus} & -- & --	&	41 \\
\bottomrule	
\end{tabular}
}
\quad
\subtable[MCLUST]
{
\label{tab:MCLUST}
\begin{tabular}{cccc}
\toprule
\backslashbox{TRUE}{Est.} 	&	 1	&	 2	\\
\midrule															
\textit{ochrogaster}  & 43	&		\textbf{2} \\
\textit{californicus} & --  &	 41 \\
\bottomrule	
\end{tabular}
}
\end{table}
Furthermore, \tablename~\ref{tab:PGMM} and \tablename~\ref{tab:MCLUST} show, respectively, the clustering results of the following model-based clustering approaches applied to the vector $\left(\bX',Y\right)'$:
\begin{description}
	\item[PGMM:] parsimonious latent Gaussian mixture models as described in \citet{McNi:Murp:Pars:2008}, \citet{McNi:Mode:2010}, and \citet{McNi:Murp:McDa:Fros:Seri:2010}, and estimated via the \texttt{pgmmEM} function of the {\sf R}-package \texttt{pgmm} (\citealp{McNi:Murp:Jamp:McDa:Bank:pgmm:2011}); and
	\item[MCLUST:] parsimonious mixtures of Gaussian distributions as described in \citet{Banf:Raft:mode:1993}, \cite{Cele:Gova:Gaus:1995}, and \citet{Fral:Raft:Mode:2002}, and estimated via the \texttt{Mclust} function of the {\sf R}-package \texttt{mclust} \citep[see][for details]{Fral:Raft:Murp:Scru:mclu:2012}.	 
\end{description}
As seen from \tablename~\ref{tab:f.voles clustering}, \textit{M.~californicus} was classified correctly using all three approaches. 
Also, \textit{M.~ochrogaster} was classified into two sub-clusters using CWFA and PGMM while MCLUST classified it into one cluster. 
However, the CWFA approach had no misclassifications between the two species but both PGMM and MCLUST misclassified two \textit{M.~ochrogaster} as  \textit{M.~californicus}.  

\begin{figure}[!ht]
\begin{center}
\resizebox{0.99\textwidth}{!}{\includegraphics[angle=-90]{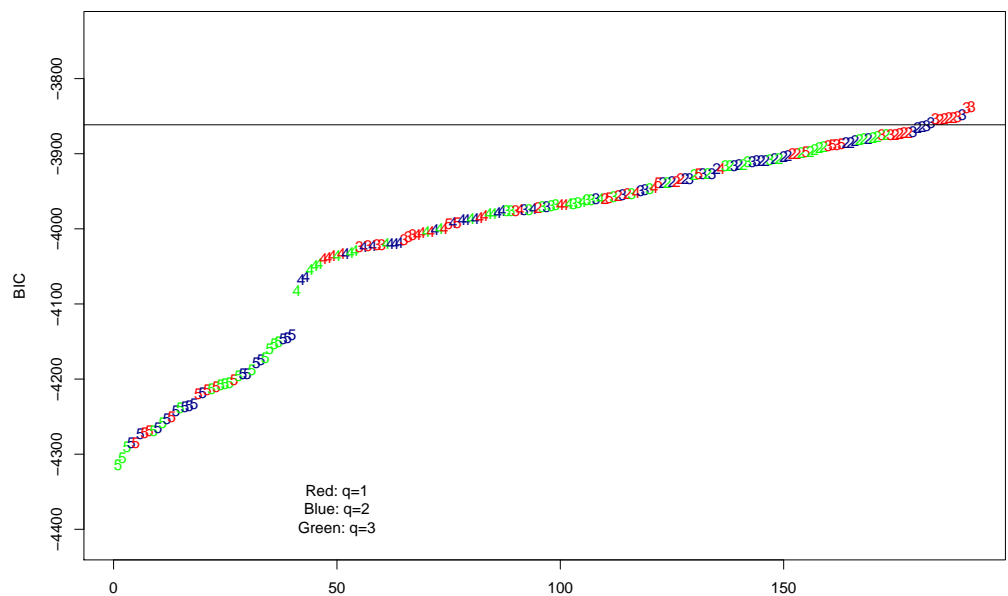}}
\caption{Group-factor plot of BIC values sorted in an increasing order for the \texttt{f.voles} data. 
Numbers denote the selected number $G$ of groups and colors denote the number $q$ of latent factors (red: $q=1$, blue: $q=2$, green: $q=3$).}
\label{fig:bicfvoles}
\end{center}
\end{figure}
\begin{figure}[!ht]
\begin{center}
\resizebox{0.9\textwidth}{!}{\includegraphics[angle=-90]{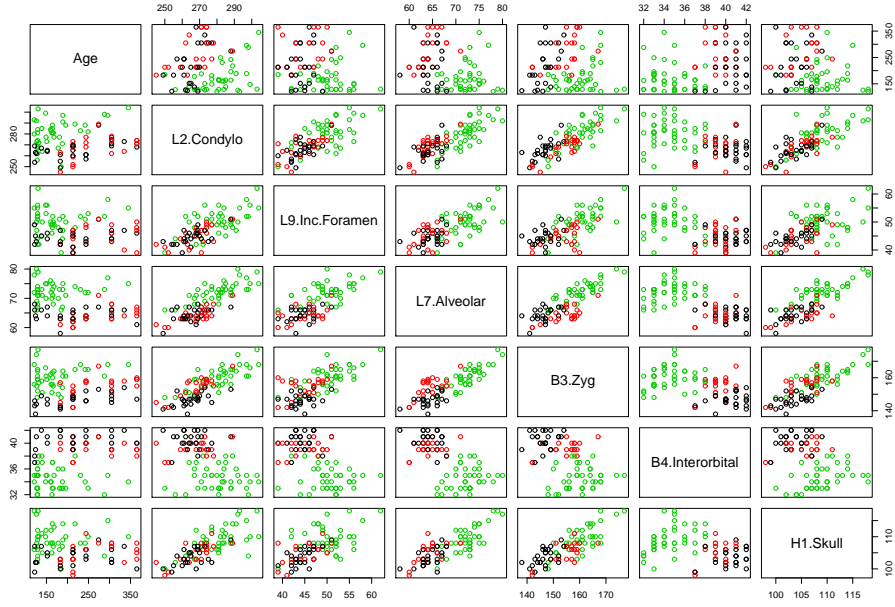}}
\caption{Scatterplot matrix of \texttt{f.voles} data showing the classification observed from CWFA modelling using the clustering framework, where black and red indicates sub-clusters of \textit{M.\ ochrogaster} species and green indicates \textit{M.\ californicus} species.}\label{fig:scatterfvoles1}
\end{center}
\end{figure}

Now, we evaluate the group-factor plot of BIC values, for all 192 models, displayed in \figurename~\ref{fig:bicfvoles}. Ten models had a BIC above the 1\% line; among them, six were characterized by $G=3$ components and the remaining four by $G=2$. 
However, from \figurename~\ref{fig:bicfvoles}, the top four models all had three components, which shows that a three component model was not randomly chosen.
\citet{Airo:Hoff:Acom:1984} mention that some unexplained geographic variation may exist among the voles. 
However, no covariate was available with such information. 
Hence, we opted for the scatter plot matrix to evaluate the presence of sub-clusters, see  \figurename~\ref{fig:scatterfvoles1}.
Here, the scatter plot of the variables $\mathsf{B}_3$ versus $\mathsf{B}_4$ shows the presence of distinct sub-clusters for \textit{M. ochrogaster}, which supports our results attained using CWFA modelling.

\subsubsection{Classification}
\label{subsubsec:Classification of real data}

A subset of observations, consisting of 50\% of the data, was randomly selected and these observations were assumed to have a known group membership. 
To allow for the unobserved sub-cluster noted in the clustering application of Section~\ref{subsubsec:Clustering on real data}, we ran the algorithm for $G=2,3$ and $q=1,2,3$. The best model (CCUU with $G=2$ and $q=1$) selected by the BIC ($-3843.482$) gave a perfect classification, as we can see from \tablename~\ref{tab:2G}.
\begin{table}[!ht]
\caption{
Classification of \texttt{f.voles} data assuming that 50\% of the observations have known group membership.
\label{tab:f.voles classification}
}
\centering
\subtable[2 known groups]{\label{tab:2G}
\begin{tabular}{ccc}
\toprule
\backslashbox{TRUE}{Est.}     	&	 1	&	 2	\\
\midrule															
\textit{ochrogaster}  & 45	&	-- \\
\textit{californicus} & -- &	41 \\
\bottomrule	
\end{tabular}
}
\quad
\subtable[3 known groups]{\label{tab:3G}
\begin{tabular}{cccc}
\toprule
\backslashbox{TRUE}{Est.}     	&	 1	&	 2	&	 3 \\
\midrule															
\textit{ochrogaster}  & 28	& 17	&	-- \\
\textit{californicus} & --&	--	&	41 \\
\bottomrule	
\end{tabular}
}
\end{table}

We also ran the classification assuming that the data are actually comprised of three known groups. Therefore, using the classification observed by clustering, we also ran the classification algorithm with 50\% known (i.e., labelled) and 50\% unknown (i.e., unlabelled). 
To further allow for the unobserved sub-cluster, we ran the algorithm for $G\in\{3,4\}$ and $q\in\{1,2,3\}$. 
The model selected using the BIC was CCCU with $G=3$ and $q=1$, with a BIC value of $-3837.383$. 
Even though the BIC value observed using the classification approach (with three known groups membership) was very close to the BIC value using clustering, the sub-clusters do not have precisely the same classification using the classification and clustering approaches. 
This could be a consequence of the classification of borderline observations among the sub-clusters using maximum \textit{a posteriori} probability.  
However, the BIC value for the classification using three known groups was higher than the BIC value using two known groups, which again suggests the presence of sub-clusters.

\section{Conclusions, discussion, and future work}
\label{sec:Conclusions, discussion, and future works}

In this paper, we introduced a novel family of 16 parsimonious mixture models for model-based clustering and classification.
They are linear Gaussian cluster-weighted models in which a latent factor structure is assumed for the explanatory random vector in each mixture component.
The parsimonious versions are obtained by combining all of the constraints described in \citet{McNi:Murp:Pars:2008} with one of the constraints illustrated in \citet*{Ingr:Mino:Punz:Mode:2012}.
Due to the introduction of a latent factor structure, the parameters are linear in dimensionality as opposed to the traditional linear Gaussian CWM where the parameters grow quadratically; therefore, our approach is more suitable for modelling complex high dimensional data. 
The AECM algorithm \citep{Meng:VanD:TheE:1997} was used for maximum likelihood estimation of the model parameters.
Being based on the EM algorithm, it is very sensitive to the starting values due to presence of multiple local maxima in high dimensional space.  
To overcome this problem, we proposed a 5-step hierarchical initialization procedure that utilizes the nested structures of the models within the family. 
Because these models have a hierarchical/nested structure, this initialization procedure guarantees a natural ranking on the likelihoods of the models in our family. 
In other words, the procedure restricts a model~A, which is nested in a model~B, from having a greater likelihood than model~B. 
Using artificial and real data, we demonstrated that these models give very good clustering performance and that the AECM algorithms used were able to recover the parameters very well. 

With regard to the latent factor structure, for a latent dimension $q>1$, the loading matrix $\boldsymbol{\Lambda}$ is unidentifiable because the model is still satisfied even when the latent factor $\bu_i$ is replaced by $\mathbf{H}u_i$ and $\boldsymbol{\Lambda}$ by $\boldsymbol{\Lambda}\mathbf{H}'$, where $\mathbf{H}$ is any orthogonal matrix of order $q$ \citep{McLa:Peel:fini:2000}. This results in an infinite number of possibilities for $\boldsymbol{\Lambda}$. Even though this does not affect the clustering algorithm interpretation of the estimated $\boldsymbol{\Lambda}$ is not informative because $\boldsymbol{\Lambda}\boldsymbol{\Lambda}'$ is unique. A future avenue of research is to explore further constraints on the factor loading matrix to ensure a uniquely defined factor loading matrix~$\boldsymbol{\Lambda}$.

Also, while the BIC was able to identify the correct model, the choice of a convenient model selection criterion for these models is still an open question. 
Some future work will be devoted to the search for good model selection criteria for these models.
Finally, here we assumed that the number of factors was the same across groups, which might be too restrictive. However, assuming otherwise also increases the number of models that need to be fitted, resulting in huge computational burden. Approaches such as variational Bayes approximations might be useful for significantly reducing the number of models that need to be fitted.

\appendix

\section{The conditional distribution of $Y|\bx,\bu$}
\label{app:Conditional distribution of Y|X,u}

To compute the distribution of $Y|\bx,\bu$, we begin by recalling that if $\bZ \sim N_q\left(\bm,\bGamma\right)$ is a random vector with values in $\real^q$ and if $\bZ$ is partitioned as $\bZ=\left(\bZ'_1,\bZ'_2\right)'$, where $\bZ_1$ takes values in
$\real^{q_1}$ and $\bZ_2$ in $\real^{q_2}=\real^{q-q_1}$, then we can write
\begin{equation*}
 \bm = 
 \begin{bmatrix} 
 \bm_1 \\ 
 \bm_2 
 \end{bmatrix} 
 \quad \text{and} \quad 
 \bGamma = 
 \begin{bmatrix} 
 \bGamma_{11} & \bGamma_{12} \\
 \bGamma_{21} & \bGamma_{22} 
 \end{bmatrix}. 
\label{Zdecomp}
\end{equation*}
Now, because $\bZ$ has a multivariate normal distribution, $\bZ_1|\bZ_2=\bz_2$ and $\bZ_2$ are statistically independent with
$\bZ_1|\bZ_2=\bz_2 \sim N_{q_1}\left(\bm_{1|2}, \bGamma_{1|2}\right)$ and $\bZ_2 \sim N_{q_2}\left(\bm_2, \bGamma_{22}\right)$,
where 
\begin{equation}
\bm_{1|2}  = \bm_1+ \bGamma_{12}\bGamma_{22}^{-1}(\bz_2-\bm_2) \quad \text{and} \quad \bGamma_{1|2}  =\bGamma_{11}-\bGamma_{12}\bGamma_{22}^{-1}\bGamma_{21}.  
\label{eq:Npar2|1}
\end{equation}
Therefore, setting $\bZ=\left(\bZ'_1,\bZ'_2\right)'$, where $\bZ'_1=Y$ and $\bZ_2=\left(\bX', \bU'\right)'$, gives $\bm_1= \beta_0 + \bbeta'_1 \bmu$ and $\bm_2 = \left(\bmu',\bzero'\right)'$, with the elements in $\bGamma$ given by
\begin{align*}
\bGamma_{11} = \bbeta'_1 \bSigma \bbeta_1 + \sigma^2,
\quad 
\bGamma_{22} & = \begin{bmatrix}  
\bSigma & \bLambda  \\
\bLambda' & \bI_q  \\
\end{bmatrix}, 
\quad \text{and} \quad \bGamma_{12} = \begin{bmatrix} \bbeta'_1 \bSigma &  \bbeta'_1\bLambda 
\end{bmatrix}.
\end{align*}
It follows that $Y|\bx,\bu$ is Gaussian with mean $\bm_{y|\bx,\bu}= \mathbb{E}\left(Y|\bx,\bu\right)$ and variance  $\sigma^2_{y|\bx,\bu}=\Var\left(Y|\bx,\bu\right)$, in accordance with the formulae in \eqref{eq:Npar2|1}. 
Because the inverse matrix of $\bGamma_{22}$ is required in \eqref{eq:Npar2|1}, the following formula for the inverse of a partitioned matrix is utilized: 
\[
\begin{bmatrix} 
\bA & \bB \\ 
\bC & \bD 
\end{bmatrix}^{-1} 
= 
\begin{bmatrix} \left(\bA - \bB \bD^{-1} \bC\right)^{-1} & -\bA^{-1} \bB \left(\bD -\bC \bA^{-1} \bB\right)^{-1} \\
-\bD^{-1} \bC \left(\bA - \bB \bD^{-1}\bC\right)^{-1} & \left(\bD -\bC\bA^{-1} \bB\right)^{-1} 
\end{bmatrix}.
\]
Again, writing $\bSigma=\bLambda \bLambda' + \bPsi$, we have
\begin{displaymath}
\bGamma_{22}^{-1} = 
\begin{bmatrix}  
\bSigma & \bLambda  \\
\bLambda' & \bI_q  \\
\end{bmatrix}^{-1} 
= 
\begin{bmatrix} 
\bPsi^{-1} & -\bSigma^{-1} \bLambda \left(\bI_q -\bLambda' \bSigma^{-1} \bLambda\right)^{-1} \\
-\bLambda' \bPsi^{-1} & \left(\bI_q -\bLambda'\bSigma^{-1} \bLambda\right)^{-1} 
\end{bmatrix}.
\end{displaymath}
Moreover, according to the Woodbury identity \citep{Wood:Inve:1950}:
$$
\bSigma^{-1}= (\bLambda \bLambda' + \bPsi )^{-1} = \bPsi^{-1} - \bPsi^{-1} \bLambda(\bI_q + \Lambda' \bPsi^{-1} \bLambda)^{-1} \bLambda'\bPsi^{-1}.
$$ 
Now,
\begin{displaymath}
\bGamma_{12}\bGamma_{22}^{-1} = 
\begin{bmatrix}  
\bbeta'_1\bSigma & \bbeta'_1\bLambda
\end{bmatrix}
\begin{bmatrix} 
\bPsi^{-1} & -\bSigma^{-1} \bLambda \left(\bI_q -\bLambda' \bSigma^{-1} \bLambda\right)^{-1} \\
-\bLambda \bPsi^{-1} & (\bI_q -\bLambda'\bSigma^{-1} \bLambda)^{-1} 
\end{bmatrix}
= \begin{bmatrix}
\bbeta'_1&0
 \end{bmatrix}.
\end{displaymath}
Finally, according to \eqref{eq:Npar2|1}, we have
\begin{eqnarray*}
\bm_{y|\bx,\bu}  & = & \bm_1+\bGamma_{12}\bGamma_{22}^{-1} 
\begin{bmatrix} 
\bz_2-\bm_2 
\end{bmatrix} 
= \left(\beta_0 + \bbeta'_1 \bmu\right) +  
\begin{bmatrix}
\bbeta'_1&0 
\end{bmatrix} 
\begin{bmatrix} 
\bx-\bmu \\ 
\bu-\bzero 
\end{bmatrix} =  
\beta_0 + \bbeta'_1 \bx, \\
\sigma^2_{y|\bx,\bu} & = & \bGamma_{11}-\bGamma_{12}\bGamma_{22}^{-1}\bGamma_{21} =\bbeta'_1\bSigma\bbeta_1+\sigma^2-
\begin{bmatrix}
\bbeta'_1&0 
\end{bmatrix}  
\begin{bmatrix} 
\bSigma \bbeta_1 \\ 
\bLambda \bbeta_1
\end{bmatrix} = \sigma^2.
\end{eqnarray*}

\section{Details on the AECM algorithm for the parsimonious models}
\label{app:details on the EM algorithm}

This appendix details the AECM algorithm for of all the models summarized in \tablename~\ref{tab:parsimonious_model}.

\subsection{Constraint on the $Y$ variable}
\label{subsec:Constraint on the Y variable}

In all of the models whose identifier starts with `C', that is the models in which the error variance terms $\sigma_g^2$ (of the response variable $Y$) are constrained to be equal across groups, i.e., $\sigma_g^2 =\sigma^2$ for $g=1,\ldots,G$, the common variance $\sigma^2$ at the $\left(k+1\right)$th iteration of the algorithm is computed as
\begin{displaymath}
\sigma^{2\left(k+1\right)} = \frac{1}{n} \sum_{i=1}^n \sum_{g=1}^G  z_{ig}^{\left(k+1\right)} \left\{ y_i-\left(\beta_{0g}^{\left(k+1\right)} + \bbeta_{1g}^{'\left(k+1\right)}\bx_i\right)\right\}^2.  
\end{displaymath}

\subsection{Constraints on the $\bX$ variable}
\label{subsec:Constraints on the X variable}

With respect to the $\bX$ variable, as explained in Section~\ref{sec:Parsimonious versions of the model}, we considered the following constraints on $\bSigma_g=\bLambda_g\bLambda_g'+\bPsi_g$: 
{\em i}) equal loading matrices $\bLambda_g = \bLambda$,
{\em ii}) equal error variance  $\bPsi_g = \bPsi$, and 
{\em iii}) isotropic assumption: $\bPsi_g = \psi_g \bI_p$.
In such cases, the $g$th term of the expected complete-data log-likelihood $Q_2\left(\btheta_2; \btheta^{(k+1/2)}\right)$, and then the estimates \eqref{update_Lambdag} and \eqref{update_Psig} in Section~\ref{sec:second}, are computed as follows.

\subsubsection{Isotropic assumption: $\bPsi_g=\psi_g\bI_p$} 

In this case, Equation \eqref{eq:expected complete-data log-likelihood2} becomes
\begin{eqnarray*}
Q_2\left(\bLambda_g,\psi_g;\btheta^{(k+1/2)}\right) & = & \text{C}\left(\btheta_1^{\left(k+1\right)}\right) +\frac{1}{2} n_g^{\left(k+1\right)} \ln |\psi^{-1}_g\bI_p | - \frac{1}{2} n_g^{\left(k+1\right)} \psi_g^{-1}\tr\left\{ \bS_g^{\left(k+1\right)}  \right\} \nonumber\\ 
&&  + 
n_g^{\left(k+1\right)} \psi^{-1}_g \text{tr}\left\{ \bgamma_g^{\left(k\right)}\bS_g^{\left(k+1\right)} \bLambda_g\right\} 
 -\frac{1}{2} n_g^{\left(k+1\right)}\psi_g^{-1}\text{tr}\left\{\bLambda_g \bTheta^{\left(k\right)}_g \bLambda_g' \right\}, 
\end{eqnarray*}
yielding
\begin{displaymath}
\frac{\partial Q_2}{\partial \psi^{-1}_g}  = \frac{1}{2} n_g^{\left(k+1\right)} \left[p\psi_g-   \tr\left\{ \bS_g^{\left(k+1\right)}  \right\}  + 2\text{tr}\left\{ \bgamma_g^{\left(k\right)}\bS_g^{\left(k+1\right)} \bLambda_g\right\}  -  \text{tr}\left\{\bLambda_g \bTheta^{\left(k\right)}_g \bLambda_g' \right\} \right]. 
\end{displaymath}
Then the estimated $\psi_g$ is attained for $\hat{\psi}_g$, satisfying
\begin{equation*}
 \frac{\partial Q_2}{\partial \psi^{-1}_g} = 0 \qquad \Rightarrow \qquad p \psi_g-   \tr\left\{ \bS_g^{\left(k+1\right)}  \right\}  + 2\text{tr}\left\{ \bgamma_g^{\left(k\right)}\bS_g^{\left(k+1\right)} \bLambda_g\right\} - \text{tr}\left\{\bLambda_g \bTheta^{\left(k\right)}_g \bLambda_g' \right\} = 0 . 
\end{equation*}
Thus, according to \eqref{update_Lambdag}, for $\bLambda_g=\hat{\bLambda}_g = \bS^{\left(k+1\right)}_g \bgamma'^{\left(k\right)}_g\bTheta_g^{-1}$ we get $\text{tr}\left\{\bLambda_g \bTheta^{\left(k\right)}_g \bLambda_g' \right\}
= \text{tr} \left\{ \bgamma_g^{\left(k\right)}\bS_g^{\left(k+1\right)} \bLambda_g \right\}$
and, finally, 
\begin{displaymath}
  \hat{\psi}_g =\frac{1}{p}\text{tr}\left\{{\bS_g^{\left(k+1\right)} -\hat{\bLambda}_g}\bgamma_g^{\left(k\right)}  \bS_g^{\left(k+1\right)}\right\}.  
\end{displaymath}
Thus,
\begin{eqnarray}
\psi^+_g & = & \frac{1}{p}\text{tr}\left\{{\bS_g^{\left(k+1\right)} - \bLambda_g}\bgamma_g^+  \bS_g^{\left(k+1\right)}\right\} 
\label{psi^+_g} 
\\
\bgamma_g^+ & = & \bLambda'_g\left(\bLambda_g\bLambda'_g+\psi_g^+ \bI_p\right)^{-1},
\nonumber 
\end{eqnarray}
with $\bTheta_g^+$ computed according to \eqref{Theta^+_g}.

\subsubsection{Equal error variance: $\bPsi_g =\bPsi$} 

In this case, from Equation~\eqref{eq:expected complete-data log-likelihood2}, we have
\begin{equation*}\begin{split}
Q_2\left(\bLambda_g, \bPsi; \btheta^{(k+1/2)}\right) &= \text{C}(\btheta_1^{\left(k+1\right)}) - \frac{1}{2} n_g^{\left(k+1\right)}\ln | \bPsi| - \frac{1}{2} n_g^{\left(k+1\right)} \tr\left\{ \bS_g^{\left(k+1\right)} \bPsi^{-1} \right\} \nonumber\\
& + 
n_g^{\left(k+1\right)}\text{tr}\left\{\bLambda_g \bgamma_g^{\left(k\right)}\bS_g^{\left(k+1\right)} \bPsi^{-1} \right\} -\frac{1}{2} n_g^{\left(k+1\right)}\text{tr}\left\{\bLambda_g'\bPsi^{-1} \bLambda_g \bTheta^{\left(k\right)}_g\right\} ,
\end{split}\end{equation*}
yielding
\begin{displaymath}
 \frac{\partial Q_2\left(\bLambda_g, \bPsi; \btheta^{(k+1/2)}\right)}{\partial \bPsi^{-1}}  = \frac{1}{2}n^{\left(k+1\right)}_g \bPsi- \frac{1}{2}n^{\left(k+1\right)}_g \bS_g^{\left(k+1\right)}+ n^{\left(k+1\right)}_g \bS_g^{'\left(k+1\right)}  \bgamma_g^{'\left(k\right)}\bLambda_g'- \frac{1}{2}n^{\left(k+1\right)}_g \bLambda_g\bTheta^{\left(k\right)}_g\bLambda_g' .
 \end{displaymath}
Then the estimated $\hat{\bPsi}$ is obtained by satisfying
\begin{equation*}
\sum_{g=1}^G \frac{\partial Q_2\left(\bLambda_g, \bPsi; \btheta^{(k+1/2)}\right)}{\partial \bPsi^{-1}}  = \bzero,
\end{equation*}
that is 
\begin{displaymath}
\frac{n}{2}\bPsi  - \frac{1}{2} \sum_{g=1}^G n^{\left(k+1\right)}_g \bS_g^{\left(k+1\right)}+
\sum_{g=1}^G  n^{\left(k+1\right)}_g \bS_g^{'\left(k+1\right)}  \bgamma_g^{'\left(k\right)}\bLambda_g'- \frac{1}{2} \sum_{g=1}^G n^{\left(k+1\right)}_g \bLambda_g\bTheta^{\left(k\right)}_g\bLambda_g' = \bzero,
\end{displaymath}
which can be simplified as
\begin{displaymath}
\frac{n}{2}\bPsi  - \frac{1}{2} \sum_{g=1}^G n^{\left(k+1\right)}_g 
\left[\bS_g^{\left(k+1\right)}+ 2 \bS_g^{'\left(k+1\right)}  \bgamma_g^{'\left(k\right)}\bLambda_g'- \bLambda_g\bTheta^{\left(k\right)}_g\bLambda_g' \right] = \bzero , 
\end{displaymath}
with $\displaystyle\sum_{g=1}^G  n^{\left(k+1\right)}_g =n$. 
Again, according to \eqref{update_Lambdag}, for $\bLambda_g=\hat{\bLambda}_g = \bS^{\left(k+1\right)}_g \bgamma'^{\left(k\right)}_g\bTheta_g^{-1}$ we get
$\hat{\bLambda}_g\bTheta^{\left(k\right)}_g \hat{\bLambda}_g' =\hat{\bLambda}_g \bgamma_g^{\left(k\right)}\bS_g^{\left(k+1\right)} $ and, afterwards,
 \begin{align}
\hat{\bPsi} &=\sum_{g=1}^G\frac{n_g}{n}\text{diag}\left\{{\bS_g^{\left(k+1\right)}-\hat{\bLambda}_g} \bgamma_g^{'\left(k\right)} \bS_g^{\left(k+1\right)}\right\} =\sum_{g=1}^G\pi_g^{\left(k+1\right)}\text{diag}\left\{{\bS_g^{\left(k+1\right)}-\hat{\bLambda}_g}\bgamma_g^{\left(k\right)} \bS_g^{\left(k+1\right)}\right\} .
\label{update_Psi}
\end{align}
Thus,
\begin{eqnarray}
\bPsi^+ & = & \sum_{g=1}^G \pi_g^{\left(k+1\right)} \text{diag}\left\{{\bS_g^{\left(k+1\right)} - \bLambda^+_g} \bgamma_g  \bS_g^{\left(k+1\right)}\right\},
\label{Psi+} 
\\ 
\bgamma_g^+ & = & \bLambda'_g\left(\bLambda^+_g \bLambda'^+_g + \bPsi^+\right)^{-1} 
\nonumber 
\end{eqnarray}
where $\bTheta_g^+$ is computed according to \eqref{Theta^+_g}.

\subsubsection{Equal loading matrices: $\bLambda_g=\bLambda$} 

In this case, Equation \eqref{eq:expected complete-data log-likelihood2} can be written as
\begin{equation*}\begin{split}
Q_2\left(\bLambda, \bPsi_g; \btheta^{(k+1/2)}\right) &=\mbox{ C}(\btheta_1^{\left(k+1\right)}) +\frac{1}{2} n_g^{\left(k+1\right)}\ln | \bPsi^{-1}_g| - \frac{1}{2} n_g^{\left(k+1\right)} \tr\left\{ \bS_g^{\left(k+1\right)} \bPsi_g^{-1} \right\} 
\\ & + n_g^{\left(k+1\right)}\text{tr}\left\{\bLambda \bgamma_g^{\left(k\right)}\bS_g^{\left(k+1\right)} \bPsi^{-1}_g \right\} -\frac{1}{2} n_g^{\left(k+1\right)}\text{tr}\left\{\bLambda'\bPsi_g^{-1} \bLambda \bTheta^{\left(k\right)}_g\right\}, 
\end{split}\end{equation*}
yielding
\begin{align*}
\frac{\partial Q_2\left(\bLambda, \bPsi_g; \btheta^{(k+1/2)}\right)}{\partial \bLambda} & = n_g^{\left(k+1\right)} \bPsi_g^{-1} \bS_g^{\left(k+1\right)} \bgamma'^{\left(k\right)}_g - n_g^{\left(k+1\right)} \bPsi_g^{-1} \bLambda   \bTheta^{\left(k\right)}_g = \bzero. 
\end{align*}
Then the estimated $\hat{\bLambda}$ is obtained by solving
\begin{equation}
\sum_{g=1}^G \frac{\partial Q_2\left(\bLambda, \bPsi_g; \btheta^{(k+1/2)}\right)}{\partial \bLambda}  = \sum_{g=1}^G n_g^{\left(k+1\right)} \bPsi_g^{-1} \left[ \bS_g^{\left(k+1\right)} \bgamma_g^{'\left(k\right)} -  \bLambda   \bTheta^{\left(k\right)}_g 
\right] = \bzero , 
\label{partialQ2_Lambda}
\end{equation}
with $\bgamma^{\left(k\right)}_g  =\bLambda^{'\left(k\right)}\left(\bLambda^{\left(k\right)}\bLambda^{'\left(k\right)}+\bPsi^{\left(k\right)}_g\right)^{-1}$.
In this case, the loading matrix cannot be solved directly and must be solved in a row-by-row manner as suggested by \cite{McNi:Murp:Pars:2008}. 
Therefore, 
\begin{eqnarray}
\lambda^+_i & = & \mathbf{r}_i\left(\sum_{g=1}^G\frac{n_g}{\psi_{g\left(i\right)}} \bTheta_g  \right)^{-1} 
\label{lambda+} \\
\bgamma_g^+ & = & \bLambda'\left(\bLambda^+ \bLambda'^+ + \bPsi^+_g\right)^{-1} 
\label{gamma^+_g_=Lambda} \\
\bTheta^+_g & = & \bI_q-\bgamma^+_g \bLambda^+ + \bgamma^+_g \bS_g^{\left(k+1\right)} \bgamma'^{+}_g , \label{Theta^+_g=Lambda} 
\end{eqnarray} 
where $\lambda^+_i$ is the $i$th row of the matrix $\bLambda^+$, $\psi_{g\left(i\right)}$ is the $i$th diagonal element of $\bPsi_g$, and $\mathbf{r}_i$ represents the $i$th row of the matrix $\displaystyle\sum_{g=1}^G n_g^{\left(k+1\right)} \left(\bPsi'_g\right)^{-1} \bS_g^{\left(k+1\right)}$.

\subsubsection{Further details}

A further schematization is here given without considering the constraint on the $Y$ variable.
Thus, with reference to the model identifier, we will only refer to the last three letters.

\begin{description}
\item[Models ended by UUU:] 
no constraint is assumed. \\
\item[Models ended by UUC:] 
$\bPsi_g =\psi_g\bI_p$, where the parameter $\psi_g$ is updated according
to \eqref{psi^+_g}. \\
\item[Models ended by UCU:] 
$\bPsi_g =\bPsi$, where the matrix $\bPsi$ is updated according to \eqref{Psi+}. \\
\item[Models ended by UCC:]  
$ \bPsi_g =\psi \bI_p$. 
By combining  \eqref{psi^+_g} and \eqref{Psi+} we obtain
\begin{equation}
  \hat{\psi} =\frac{1}{p}\sum_{g=1}^G \frac{n_g^{\left(k+1\right)}}{n}\text{tr}\left\{{\bS_g^{\left(k+1\right)} -\hat{\bLambda}_g}\bgamma_g^{\left(k\right)}  \bS_g^{\left(k+1\right)}\right\} =
  \frac{1}{p}\sum_{g=1}^G \hat{\pi}_g^{\left(k+1\right)} \text{tr}\left\{{\bS_g^{\left(k+1\right)} -\hat{\bLambda}_g}\bgamma_g^{\left(k\right)}  \bS_g^{\left(k+1\right)}\right\}. 
  \label{update_UCC}
\end{equation}
Thus,
\begin{eqnarray*}
\psi^+ & = & \frac{1}{p}\sum_{g=1}^G \pi_g^{\left(k+1\right)} \text{tr}\left\{{\bS_g^{\left(k+1\right)} - \bLambda^+_g} \bgamma_g  \bS_g^{\left(k+1\right)}\right\} 
\\
\bgamma^+_g & = & \bLambda'^+_g\left(\bLambda^+_g \bLambda'^+_g+\psi^+ \bI_p\right)^{-1}, 
\end{eqnarray*}
with $\bTheta_g^+$ computed according to \eqref{Theta^+_g}. \\
\item[Models ended by CUU:]   
$\bLambda_g =\bLambda$, where the matrix $\bLambda$ is updated according to \eqref{lambda+}. 
In this case, $\bPsi_g$ is estimated directly from \eqref{Q2Psi} and thus
$\bPsi^+_g =\text{diag}\left\{\bS_g^{\left(k+1\right)}-2 \bLambda^+ \bgamma_g \bS_g^{\left(k+1\right)} +\bLambda^+ \bTheta_g \bLambda'^+ \right\}$, 
with $\bgamma_g^+$ and $\bTheta^+_g$ computed according to \eqref{gamma^+_g_=Lambda} and \eqref{Theta^+_g=Lambda}, respectively. \\
\item[Models ended by CUC:]  $\bLambda_g =\bLambda$ and $\bPsi_g =\psi_g\bI_p$.  
In this case, equation~\eqref{partialQ2_Lambda}, for $\bPsi_g =\psi_g\bI_p$, yields
\begin{equation*}
\sum_{g=1}^G \frac{\partial Q_2 \left(\bLambda, \psi_g; \btheta^{(k+1/2)}\right)}{\partial \bLambda}  =  	\sum_{g=1}^G n_g^{\left(k+1\right)} \psi_g^{-1}\bS_g^{\left(k+1\right)} \bgamma_g^{'\left(k\right)} -  \sum_{g=1}^G n_g^{\left(k+1\right)} \psi_g^{-1} 
  \bTheta^{\left(k\right)}_g  = \bzero , \,  
\end{equation*}
and afterwards
\begin{displaymath}
\hat{\bLambda} = \left( \sum_{g=1}^G \frac{n_g^{\left(k+1\right)}}{\psi_g^{-1}}  \bS_g^{\left(k+1\right)} \bgamma_g^{'\left(k\right)} \right) \left( \sum_{g=1}^G \frac{n_g^{\left(k+1\right)}}{\psi_g^{-1}} \bLambda   \right)^{-1}, 
\end{displaymath}
with $\bgamma^{\left(k\right)}_g  =\bLambda^{'\left(k\right)}\left(\bLambda^{\left(k\right)}\bLambda^{'\left(k\right)}+\psi^{\left(k\right)}_g \bI_p\right)^{-1}$. 
Moreover, from 
\begin{align*}
 \frac{\partial Q_2\left(\bLambda, \psi_g; \btheta^{(k+1/2)}\right)}{\partial \psi_g^{-1}}  &=\frac{p}{2} \psi_g -\frac{n_g^{\left(k+1\right)}}{2}  \left[\text{tr}\left\{\bS_g^{\left(k+1\right)}\right\}-2  \text{tr}\left\{\bS_g^{'\left(k+1\right)} \bgamma'^{\left(k\right)}_g \bLambda'\right\}+ \text{tr}\left\{\bLambda \bTheta_g^{\left(k+1\right)} \bLambda'\right\}\right] \\&= 0
\end{align*}
we get
$\hat{\psi}_g =({1}/{p})\text{tr}\left\{ \bS_g^{\left(k+1\right)}-2\hat{\bLambda} \bgamma'^{\left(k\right)}_g \bS_g+\hat{\bLambda} \bTheta_g\hat{\bLambda}'\right\}$. 
Thus,
\begin{eqnarray*}
\bLambda^+ & = &\left(\sum_{g=1}^G \frac{n_g^{\left(k+1\right)}}{\psi_g^{-1}}  \bS_g^{\left(k+1\right)} \bgamma_g' \right) \left( \sum_{g=1}^G \frac{n_g^{\left(k+1\right)}}{\psi_g^{-1}} \bLambda   \right)^{-1} 
\\
\psi^+_g  &  = &\frac{1}{p}\text{tr}\left\{ \bS_g^{\left(k+1\right)}-2\ \bLambda^+ \bgamma'_g \bS_g+ \bLambda^+ \bTheta \bLambda'^+ \right\} 
\\
\bgamma^+_g & = &\bLambda'^+\left(\bLambda^+\bLambda'^+ +\psi^+_g \bI_p\right)^{-1} .
\end{eqnarray*}
with $\bTheta_g^+$ computed according to \eqref{Theta^+_g=Lambda}. \\
\item[Models ended by CCU:] $\bLambda_g =\bLambda$ and $\bPsi_g =\bPsi$, so that $\bgamma^{\left(k\right)} =\bLambda'^{\left(k\right)}\left(\bLambda^{\left(k\right)}\bLambda^{\left(k\right)}+\bPsi^{\left(k\right)}\right)^{-1}$. 
Setting $\bPsi_g=\bPsi$ in \eqref{partialQ2_Lambda}, we get
\begin{align*}
\sum_{g=1}^G \frac{\partial Q_2\left(\bLambda, \bPsi; \btheta^{(k+1/2)}\right)}{\partial \bLambda}  & = \sum_{g=1}^G n_g^{\left(k+1\right)} \bPsi^{-1} \left[ \bS_g^{\left(k+1\right)} \bgamma'^{\left(k\right)} -  \bLambda   \bTheta^{\left(k\right)}_g 
\right] \\ & = \bPsi^{-1}  \left[ \bgamma'^{\left(k\right)} \sum_{g=1}^G n_g^{\left(k+1\right)} \bS_g^{\left(k+1\right)}  -  \bLambda \sum_{g=1}^G n_g^{\left(k+1\right)}  \bTheta^{\left(k\right)}_g \right] \\ 
& = \bPsi^{-1}  \left[ \bgamma'^{\left(k\right)}  \bS^{\left(k+1\right)}  -  \bLambda  \bTheta^{\left(k\right)} \right] = \bzero,
\end{align*}
where 
\begin{align*}
\bS^{\left(k+1\right)} &= \sum_{g=1}^G \pi_g^{\left(k+1\right)} \bS_g^{\left(k+1\right)}  \\
\bTheta^{\left(k\right)} & =\sum_{g=1}^G \pi_g^{\left(k+1\right)}  \bTheta^{\left(k\right)}_g  =   \bI_q-\bgamma^{\left(k\right)} \bLambda^{\left(k\right)}+ 
\bgamma^{\left(k\right)} \bS^{\left(k+1\right)} \bgamma'^{\left(k\right)}. \, 
\end{align*}
Thus,
\begin{equation}
\hat{\bLambda} =\bS^{\left(k+1\right)}  \bgamma'^{\left(k\right)}  \left(\bTheta^{\left(k\right)}\right)^{-1}. \label{update_Lambda_CCU}
\end{equation}
Moreover, setting $\bLambda_g=\bLambda$ in \eqref{update_Psi}, we get
$\hat{\bPsi} = \text{diag}\left\{{\bS^{\left(k+1\right)}-\hat{\bLambda}}\bgamma^{\left(k\right)} \bS^{\left(k+1\right)}\right\}$. 
Hence,
\begin{eqnarray}
\bLambda^+ & = & \bS^{\left(k+1\right)}  \bgamma'  \bTheta^{-1} 
\label{Lambda+_CCU} 
\\
\bPsi^+  & = & \text{diag}\left\{ \bS^{\left(k+1\right)}- \bLambda^+ \bgamma \bS^{\left(k+1\right)}\right\}  
\nonumber
\\
\bgamma^+_g & = & \bLambda'^+\left(\bLambda^+\bLambda'^+ +\bPsi^+\right)^{-1}, 
\nonumber
\end{eqnarray}
with $\bTheta_g^+$ computed according to \eqref{Theta^+_g=Lambda}. \\
\item[Models ended by CCC:] $\bLambda_g =\bLambda$ and $\bPsi_g =\psi \bI_p$, so that $\bgamma^{\left(k\right)} =\bLambda'^{\left(k\right)}\left(\bLambda^{\left(k\right)}\bLambda'^{\left(k\right)}+\psi^{\left(k\right)}\right)^{-1}$. 
Here, the estimated loading matrix is again \eqref{update_Lambda_CCU}, while the isotropic term obtained from \eqref{update_UCC} for $\bLambda_g=\bLambda$ is
$\hat{\psi} = ({1}/{p})\text{tr}\left\{\bS^{\left(k+1\right)} - \hat{\bLambda}\bgamma^{\left(k\right)}  \bS^{\left(k+1\right)}\right\}$, 
with $\bgamma^{\left(k\right)}_g  =\bLambda'^{\left(k\right)}_g\left(\bLambda^{\left(k\right)}_g\bLambda'^{\left(k\right)}_g+\psi^{\left(k\right)} \bI_p\right)^{-1}$.
Hence,
\begin{eqnarray*}
\psi^+ & = & \frac{1}{p} \text{tr}\left\{\bS^{\left(k+1\right)} - \bLambda^+ \bgamma   \bS^{\left(k+1\right)}\right\}  
\\
\bgamma^+ & = & \bLambda'^+\left(\bLambda^+\bLambda'^+ + \psi^+ \bI_p\right)^{-1}, 
\end{eqnarray*}
with $\bLambda^+$ and $\bTheta_g^+$ computed according to \eqref{Lambda+_CCU} and \eqref{Theta^+_g=Lambda}, respectively.
\end{description}

\section{True and estimated covariance matrices of Section~\ref{subsec:Simulated data}}
\label{app:True and estimated covariance matrices}

Because the loading matrices are not unique, for the simulated data of Examples~1 and~2 we limit the attention to a comparison, for each $g=1,\ldots,G$, of true and estimated covariance matrices.

\subsection{Example~\ref{Ex:ExampleV2}}
\label{app:Example 1} 

\begin{displaymath}
\boldsymbol{\Sigma}_1= 
\begin{bmatrix}[rrrrr]
103.36 & 103.07& 101.37&  79.41& 105.66\\
103.08 & 119.39& 110.23&  85.97& 115.47\\
101.37 & 110.23& 129.77& 106.08& 118.50\\
79.41  &  85.97& 106.08& 101.46&  95.21\\
105.66 & 115.47& 118.50&  95.21& 121.63
\end{bmatrix}
\quad
\hat{\boldsymbol{\Sigma}}_1 =
\begin{bmatrix}[rrrrr]
107.59&114.55&110.42&87.29&114.43\\
114.55&139.40& 127.06& 100.09&132.06\\
110.42& 127.06& 146.31& 122.92&134.12\\
87.29& 100.09& 122.92& 117.97&110.09\\
114.43&132.06& 134.12& 110.09& 135.66\\
\end{bmatrix},
\end{displaymath}
and
\begin{displaymath}
\boldsymbol{\Sigma}_2= 
\begin{bmatrix}[rrrrr]
 34.25&15.16& 17.81& 22.39&14.62\\
15.16& 17.01& 11.42& 13.98&  8.95\\
17.81& 11.42& 17.62& 16.12& 10.45\\
 22.39& 13.98& 16.12& 28.11& 13.11\\
14.62&  8.95& 10.45& 13.11& 10.19
\end{bmatrix}
\quad
\hat{\boldsymbol{\Sigma}}_2 =
\begin{bmatrix}[rrrrr]
 22.16&  7.44& 13.71& 12.89& 10.12\\
 7.44& 11.25&  7.59&  8.05&  5.48\\
 13.71&  7.59& 18.83& 13.53& 10.13\\
12.89&  8.05& 13.53& 22.00 & 9.41\\
10.12&  5.48& 10.13&  9.41&  8.63
\end{bmatrix}.
\end{displaymath}

\subsection{Example~\ref{Ex:ExampleV3}}
\label{appendixe2}

\begin{displaymath}
\boldsymbol{\Sigma}_1= 
\begin{bmatrix}[rrrrr]
10.41& 3.61& 4.07& 4.48&  5.71\\
3.61& 7.83& 2.88& 3.18&  4.03\\
4.07& 2.88& 8.67& 3.81&  4.64\\
4.48& 3.18& 3.81& 9.61&  5.17\\
5.71& 4.04& 4.64& 5.17& 11.73
\end{bmatrix}
\quad
\hat{\boldsymbol{\Sigma}}_1 =
\begin{bmatrix}[rrrrr]
8.86&3.89&5.06& 3.84& 5.72\\
 3.89& 7.23& 3.59&1.79& 4.04\\
 5.06& 3.59& 8.44& 3.85& 5.50\\
3.84& 1.79& 3.85& 7.74& 4.38\\
 5.72& 4.04& 5.50& 4.38& 9.81
\end{bmatrix},
\end{displaymath}
\begin{displaymath}
\boldsymbol{\Sigma}_2= 
\begin{bmatrix}[rrrrr]
103.36& 103.07& 101.37&  79.41& 105.66\\
103.08& 122.1& 110.23&  85.97& 115.47\\
101.37& 110.23& 134.33& 106.08& 118.50\\
79.41& 85.97&106.08& 102.73&95.21\\
105.66& 115.47& 118.50&  95.21&129.21
\end{bmatrix}
\quad
\hat{\boldsymbol{\Sigma}}_2 =
\begin{bmatrix}[rrrrr]
106.17&100.46&93.18&73.81&105.01\\
100.46& 113.71&  92.97& 72.22& 107.88\\
 93.18&  92.97& 108.25& 83.08& 102.36\\
73.81&  72.22&  83.08& 80.09&  81.85\\
105.01& 107.88& 102.36& 81.85& 122.59
\end{bmatrix},
\end{displaymath}
and 
\begin{displaymath}
\boldsymbol{\Sigma}_3= 
\begin{bmatrix}[rrrrr]
25.19& 15.16& 17.81& 22.39& 14.62\\
15.16& 10.67& 11.42& 13.98&  8.95\\
17.81& 11.42& 13.12& 16.12& 10.45\\
 22.39& 13.98& 16.12& 20.31& 13.11\\
14.62&  8.95& 10.45& 13.11&  8.70
\end{bmatrix}
\quad
\hat{\boldsymbol{\Sigma}}_3 =
\begin{bmatrix}[rrrrr]
32.47& 19.91& 23.06& 28.78& 18.80\\
19.91& 14.10& 14.96& 18.25& 11.66\\
23.06& 14.96& 16.95& 20.77& 13.45\\
28.78& 18.25& 20.77& 25.95& 16.77\\
18.80& 11.66& 13.45& 16.77& 11.10
\end{bmatrix}.
\end{displaymath}

\bibliographystyle{chicago}
\bibliography{References}

\end{document}